\documentclass[a4paper,11pt]{article}
\usepackage{jheppub}
\usepackage{graphicx,color,amsmath}

\title{ \boldmath Impact of  a four-zero Yukawa texture on $h\to \gamma \gamma$ and $\gamma Z$
 in the framework of the
Two Higgs Doublet Model Type III }

\author[a]{A. Cordero-Cid}
\emailAdd{acordero@ece.buap.mx}
\affiliation[a]{Fac. de Cs. de la
Electr\'onica, Benem\'erita Universidad Aut\'onoma de Puebla,
Apdo. Postal 542, 72570 Puebla, Puebla, M\'exico}

\author[b]{J. Hern\' andez--S\' anchez}
\emailAdd{jaime.hernandez@correo.buap.mx}
\affiliation[b]{Fac. de Cs. de la
Electr\'onica, Benem\'erita Universidad Aut\'onoma de Puebla,
Apdo. Postal 542, 72570 Puebla, Puebla, M\'exico and Dual C-P Institute of High Energy Physics, M\'exico.}

\author[c]{C. G. Honorato}
\affiliation[c]{Departamento de F\'{\i}sica, CINVESTAV, Apdo.
Postal 14-740, 07000 M\'exico, D. F., M\'exico.}

\author[d]{S. Moretti}
\emailAdd{s.moretti@soton.ac.uk}
\affiliation[d]{School of Physics and Astronomy, University of Southampton, Highfield, Southampton SO17 1BJ, United Kingdom, and Particle Physics Department, Rutherford Appleton Laboratory, Chilton, Didcot, Oxon OX11 0QX, United Kingdom}

\author[c]{M. A. P\'erez }
\emailAdd{mperez@fis.cinvestav.mx}
\author[e]{A. Rosado}
\emailAdd{rosado@ifuap.buap.mx}
\affiliation[e]{ Instituto de
F\'{\i}sica, BUAP, Apdo. Postal J-48, C.P. 72570 Puebla, Pue.,
M\'exico.}

\abstract{We study  the substantial enhancement, with respect to the corresponding Standard Model rates, that can be obtained
for the branching ratios of the decay channels $h \to \gamma \gamma$ and $h\to \gamma Z$ within the framework of the Two Higgs Doublet Model Type III,  assuming a four-zero Yukawa texture and a general Higgs potential.  We show that these processes are very sensitive to the flavor pattern entering the Yukawa texture and to the
 triple coupling structure of the Higgs potential, both of which impact onto the aforementioned decays. We can accommodate the parameters of the model in such a way  to obtain  the $h \to \gamma \gamma$ rates reported by the Large Hadron Collider and at the same time we get  a $h\to \gamma Z$ fraction much larger than in the Standard Model,
indeed within experimental reach.  We present some scenarios where this phenomenology is realized for spectrum configurations that are consistent with current constraints.  We also discuss the possibility of obtaining a light charged Higgs boson compatible with all such measurements, thereby serving the purpose of providing a hallmark signal of the scenario considered.}

\begin{document}
\maketitle
\flushbottom

\section{Introduction}

 New physics effects in the radiative decays of Higgs bosons
have been studied for more than twenty years \cite{Martinez:1989bg}. In particular, already
within the effective Lagrangian approach \cite{Wudka:1994ny}, it was pointed
out that anomalous contributions to the Standard Model (SM) vertices $WW \gamma$ and
$WWH$ could induce an enhancement of the Branching ratio (Br) expected
for the two-photon decay mode of the SM Higgs boson \cite{Perez:1992sa}. This topic has been the subject of renewed interest after the recent
announcement of the discovery of a new neutral scalar boson, first hinted by CDF and D0 in a wide mass interval between 115 and 130 GeV or so \cite{Aaltonen:2012qt}, then finally confirmed
with a  mass of 125.2 $\pm $0.3$\pm$0.6 GeV and 125.8$\pm$0.4$\pm$0.4 GeV
by the ATLAS and CMS Collaborations, respectively \cite{:2012gk,:2012gu}.
The new particle seen  at the Fermi National Accelerator Laboratory (FNAL) and Large Hadron Collider (LHC) is presently rather  compatible with the neutral Higgs boson of the SM
 \cite{stanmod}. However, this is not a certainty and the LHC will aim at establishing once and forever whether such an object is really  the Higgs
 particle of the SM (or not) during its upcoming runs \cite{djouadi, djouadi2}. In fact, following the initial discovery announcement on 4$^{\rm th}$ July 2012,
there has been much speculation about the excess of events in the decay channel
$h\to \gamma \gamma$ initially suggested by both ATLAS and CMS \cite{:2012gk,:2012gu}, though more recently CMS (but not ATLAS) have claimed an opposite
effect \cite{ATLASMoriond,CMSMoriond}. This potential excess could be explained by the
 existence of additional charged particles running in the loops of the radiative  Higgs coupling to photons, how it
 happens in some extended Higgs sectors
\cite{Ferreira:2012my,Kanemura:2012rs,Kitahara:2012pb,Delgado:2012sm,Chiang:2012qz,alves.2012,craig.2012,wolfgang,moretti,Basso:2013wna,Basso:2013hs, moretti0535, celis,maria,shu, Han:2012dd, Cao:2012fz,Cao:2013ur}. Conversely, if this enhancement in
$ h\to \gamma \gamma$ disappears, it will still constrain the parameter
 space of various extensions of the SM. Another  decay channel that is closely related to the di-photon one and that might give another
 clean signal in the LHC  experiments is the $\gamma Z$ mode, wherein the same new charged particles  would  contribute \cite{Djouadi:1996yq}.
Despite being  highly suppressed processes, the $h \to \gamma \gamma$ and $h \to \gamma Z$ decays, for the above reason, can nonetheless  offer a window of
understanding into possible Beyond the SM (BSM) scenarios even when no new states are found in real processes. In particular, the simultaneous measurement of
these channels at the LHC will (eventually) provide us with significant information about the possible underlying structure of the Higgs sector, as in most BSM scenarios
the rates of these two channels scale (almost identically, in most cases) with respect to the SM ones
\cite{Carena:2012xa,Chiang:2012qz,Chen:2013vi}. The upcoming higher-energy LHC run, which is expected to start in 2015 at $\sqrt{s} \approx 13 - 14$ TeV with 100 fb$^{-1}$ of luminosity per year,  will greatly
extend the experimental sensitivity to BSM physics, irrespectively of whether it is produced through real or virtual dynamics. Furthermore, one of many currently discussed
$e^+e^-$ facilities,
like the International Linear Collider (ILC) \cite{Djouadi:2007ik}, the Compact Linear Collider (CLIC) \cite{Linssen:2012hp} and the Triple Large Electron-Positron (TLEP) collider
\cite{Gomez-Ceballos:2013zzn}, may be commissioned within a decade or so, thereby offering the possibility of carrying out high precision Higgs analyses. Therefore,
it is very timely to study the scope of the $\gamma\gamma$ and $\gamma Z$ signatures in disentangling a possible non-minimal structure of the Higgs sector.

In this paper, we address the potential, in the above respect, of
 the most general version of a Two Higgs Doublet Model which is of Type III (2HDM-III),
wherein the fermionic couplings of the ensuing neutral scalars are non-diagonal in flavor
and the Higgs potential is the most general one compatible with Electro-Weak Symmetry Breaking (EWSB) (and CP conservation).
This framework, however, potentially embeds unwanted Flavor Changing Neutral Current (FCNC) phenomena \cite{Branco:2011iw}.
The simplest and most common approach to avoid these is to
impose a $\mathcal{Z}_2$ symmetry forbidding all non-diagonal  terms in flavor space
in the model Lagrangian \cite{Glashow:1976nt}. Herein, we focus instead on the
version where the Yukawa couplings depend on the
hierarchy of masses. This construct is the one where the mass matrix has a four-zero
texture form \cite{fritzsch} forcing the non-diagonal Yukawa couplings to be proportional to the geometric mean of the two fermion masses
involved \cite{Cheng:1987rs,DiazCruz:2004pj}. This matrix is based on the phenomenological
observation that the off-diagonal elements have to be small in order to dim the interactions that violate flavor, as innumerable experimental results
show.

In the next section, we briefly describe the theoretical structure of the Yukawa sector in the 2HDM-III. In section III,
we present the Feynman rules for the $\gamma \gamma \phi$ and for $\gamma Z \phi$ interactions (where $\phi$ signifies the
intervening Higgs boson, either CP-even or CP-odd). In section IV, we present
our numerical results. In section V, we summarize and conclude. Finally, some more technical details of the calculations are relegated to
the Appendix.

\section{The  Higgs-Yukawa sector  of the 2HDM-III}

The 2HDM includes two Higgs scalar doublets of hypercharge $+1$:
$\Phi^\dag_1=(\phi^-_1,\phi_1^{0*})$ and
$\Phi^\dag_2=(\phi^-_2,\phi_2^{0*})$. The most general $SU(2)_L \times U(1)_Y $
invariant  scalar potential  can be written as~\cite{Gunion:2002zf}
\begin{eqnarray}
V(\Phi_1,\Phi_2)&=&\mu^2_1(\Phi_1^\dag
\Phi_1)+\mu^2_2(\Phi^\dag_2\Phi_2)-\left(\mu^2_{12}(\Phi^\dag_1\Phi_2)+{\rm
H.c.}\right) + \frac{1}{2}
\lambda_1(\Phi^\dag_1\Phi_1)^2 \\ \nonumber
&& +\frac{1}{2} \lambda_2(\Phi^\dag_2\Phi_2)^2+\lambda_3(\Phi_1^\dag
\Phi_1)(\Phi^\dag_2\, \Phi_2)
+\lambda_4(\Phi^\dag_1\Phi_2)(\Phi^\dag_2\Phi_1) \\ \nonumber
&& +
\left(\frac{1}{2} \lambda_5(\Phi^\dag_1\Phi_2)^2+\left(\lambda_6(\Phi_1^\dag
\Phi_1)+\lambda_7(\Phi^\dag_2\Phi_2)\right)(\Phi_1^\dag \Phi_2)+
{\rm H.c.}\right), \label{potential}
\end{eqnarray}
where all parameters are assumed to be real, including the scalar field vacuum expectation values $\langle \Phi \rangle ^\dag_1=(0,v_1)$ and
$\langle \Phi \rangle ^\dag_2=(0,v_2)$, namely, both explicit and spontaneous CP-violation do not   
occur\footnote{The  $\mu^2_{12}$, $\lambda_5$, $\lambda_6$ and $\lambda_7$ parameters are complex in general, but we will assume that they are real for simplicity.}. 
When a
specific four-zero texture is implemented as a flavor symmetry in the Yukawa sector, discrete
symmetries in the Higgs potential are not needed.
Hence, one must keep the terms proportional to $\lambda_6$ and $\lambda_7$.  These parameters play an important
role in one-loop processes though, where self-interactions
of Higgs bosons could be relevant \cite{HernandezSanchez:2011fq}. In particular, with our assumptions, the  Higgs potential  is not invariant under the so-called custodial symmetryl $SU(2)_L \times SU(2)_R$ only when 
$\lambda_4 \neq \lambda_5$ \cite{Pomarol:1993mu,Branco:2011iw}. Then, the possibility of large contributions to the $\rho = m_W^2/m_Z^2 \cos ^2 \theta_W$ parameter comes only from the difference $(\lambda_4 - \lambda_5) $, which can be rewritten in terms of $(m_{H^\pm}^2 - m_A^2)$, being large. In Ref. \cite{Gunion:2002zf}, we can get the general expression of the Higgs spectrum and one obtains in particular the squared mass for the charged Higgs state:
 \begin{eqnarray} 
m_{H^\pm}^2 = m_A^2 + \frac{1}{2} v ^2 ( \lambda_4 - \lambda_5). 
\label{mhcg}
\end{eqnarray}
  Recently, another possibility was studied in Ref. \cite{Gerard:2007kn}, where a twisted custodial symmetry is presented and generalizes the case above. This symmetry  is broken when   $m_{H^\pm} - m_A$ or  $m_{H^\pm} - m_H$ are sizable.  In both cases, we must also consider  the corresponding mass of the CP-even neutral Higgs $H$-state:
\begin{eqnarray}
m_H^2 = m_{A}^2 + v^2\bigg( \lambda -\lambda_A + \hat{\lambda} \frac{\cos (\beta -\alpha)}{\sin (\beta -\alpha)} \bigg),
\end{eqnarray}
where the parameters $\lambda$, $\lambda_A$ and $\hat{\lambda}$ are given in Ref. \cite{Gunion:2002zf} and are  functions of all parameters $\lambda_i$. Following the analysis of this reference, we can get  in the SM-like scenario ($\cos  (\beta -\alpha) \to 0$) that  $(m_{A}^2 - m_H^2)= {\cal{O}} (v^2)$  and, using eq. (\ref{mhcg}),  we can also relate    $m_{H^\pm} - m_H $ to the difference $  (\lambda_4 - \lambda_5)$.   Consequently, the parameters $\lambda _6$ and $\lambda_7$ are not so relevant in the contributions to the $\rho$ parameter. Besides, the twisted symmetry allows for a scenario where the pseudoscalar Higgs state is light \cite{Branco:2011iw,deVisscher:2009zb}, which will be  discussed below.
  As  the Higgs potential  has CP-conservation, one can avoid mixing among the real and imaginary parts of the neutral scalar fields,  so that the general expressions of  the oblique parameters  are reduced to those  given in Ref. \cite{Kanemura:2011sj}\footnote{ When the most general Higgs potential with CP-violation is considered, one must use the general expressions of the oblique parameters given in  \cite{Grimus:2007if,Grimus:2008nb}.}.
Although the parameters $\lambda_6$ and $\lambda_7 $ can avoid to be constrained by the $\rho$ parameter, there are other ways to subject  them to various tests, e.g.,  perturbativity and unitarity \cite{Branco:2011iw}.  In particular, we found that the strongest constraint   for the most general Higgs potential of the 2HDM  comes from tree-level unitarity \cite{Ginzburg:2005dt}. We found numerically  the following constraint for $\tan \beta \leq 10$:
\begin{eqnarray}
|\lambda_{6,7}| \leq 1,
\end{eqnarray}  
which will be used in all our subsequent work.

In order to derive the interactions of the type Higgs-fermion-fermion, the
Yukawa Lagrangian is written as follows:
{\small
\begin{equation}
 {\cal{L}}_{Y}  = -\Bigg(
Y^{u}_1\bar{Q}_L {\tilde \Phi_{1}} u_{R} +
                   Y^{u}_2 \bar{Q}_L {\tilde \Phi_{2}} u_{R} +
Y^{d}_1\bar{Q}_L \Phi_{1} d_{R}
 + Y^{d}_2 \bar{Q}_L\Phi_{2}d_{R} +Y^{{l}}_{1}\bar{L_{L}}\Phi_{1}l_{R} +Y^{{l}}_{2}\bar{L_{L}}\Phi_{2}l_{R} \Bigg),
\label{lag-f}
\end{equation} }
\noindent where $\Phi_{1,2}=(\phi^+_{1,2},
\phi^0_{1,2})^T$ refer to the two Higgs doublets, ${\tilde
\Phi_{1,2}}=i \sigma_{2}\Phi_{1,2}^* $.
After spontaneous EWSB, one can derive the
fermion mass matrices from eq. (\ref{lag-f}), namely:
$ M_f= \frac{1}{\sqrt{2}}(v_{1}Y_{1}^{f}+v_{2}Y_{2}^{f})$, $f = u$, $d$, $l$,
assuming that both Yukawa matrices $Y^f_1$ and $Y^f_2$ have the
four-texture form and are Hermitian \cite{DiazCruz:2004pj}.
 The diagonalisation is
performed in the following way:
$ \bar{M}_f = V_{fL}^{\dagger}M_{f}V_{fR}$. Then,
$\bar{M}_f=\frac{1}{\sqrt{2}}(v_{1}\tilde{Y}_{1}^{f}+v_{2}
\tilde{Y}_{2}^{f})$,
where $\tilde{Y}_{i}^{f}=V_{fL}^{\dagger}Y_{i}^{f}V_{fR}$.
One can derive a better approximation for the product
$V_q\, Y^{q}_n \, V_q^\dagger$, by expressing the
rotated matrix $\tilde {Y}^q_n$ as
\begin{eqnarray}
\left[ \tilde{Y}_n^{q} \right]_{ij}
= \frac{\sqrt{m^q_i m^q_j}}{v} \, \left[\tilde{\chi}_{n}^q \right]_{ij}
=\frac{\sqrt{m^q_i m^q_j}}{v}\,\left[\chi_{n}^q \right]_{ij}  \, e^{i \vartheta^q_{ij}},
\label{cheng-sher}
\end{eqnarray}
\noindent
where the $\chi$'s are unknown dimensionless parameters of the model.
Following the recent analysis of \cite{HernandezSanchez:2012eg} (see also \cite{Felix-Beltran:2013tra}), we can obtain the generic expression
for the interactions of  the Higgs bosons with the fermions,
{\small
\begin{eqnarray}
{\cal L}^{\bar{f}_i f_j \phi}  & = &
-\left\{\frac{\sqrt2}{v}\overline{u}_i
\left(m_{d_j} X_{ij} {P}_R+m_{u_i} Y_{ij} {P}_L\right)d_j \,H^+
+\frac{\sqrt2m_{{l}_j} }{v} Z_{ij}\overline{\nu_L^{}}{l}_R^{}H^+
+{H.c.}\right\} \nonumber \\
& &-
\frac{1}{v} \bigg\{ \bar{f}_i m_{f_i} h_{ij}^f  f_j h^0 + \bar{f}_i m_{f_i} H_{ij}^f  f_j H^0 - i \bar{f}_i m_{f_i} A_{ij}^f  f_j \gamma_5 A^0\bigg\},
\label{lagrangian-f}
\end{eqnarray}
where $\phi_{ij}^f$ ($\phi=h$, $H$, $A$), $X_{ij}$, $Y_{ij}$ and $Z_{i j}$ are defined as follows:
\begin{eqnarray}\label{hHA}
\phi_{ij}^f & = & \xi_\phi^f \delta_{ij} + G(\xi_\phi^f,X), \, \, \, \phi= h, H, A,    \\
X_{i j} & = &   \sum^3_{l=1}  (V_{\rm CKM})_{il} \bigg[ X \, \frac{m_{d_{l}}}{m_{d_j}} \, \delta_{lj}
-\frac{f(X)}{\sqrt{2} }  \,\sqrt{\frac{m_{d_l}}{ m_{d_j} }} \, \tilde{\chi}^d_{lj}  \bigg],
 \label{Xij} \\
Y_{i j} & = &  \sum^3_{l=1}  \bigg[ Y  \, \delta_{il}
  -\frac{f(Y)}{\sqrt{2} }  \,\sqrt{\frac{ m_{u_l}}{m_{u_i}} } \, \tilde{\chi}^u_{il}  \bigg]  (V_{\rm CKM})_{lj},
\\
Z_{i j}^{l}& = &   \bigg[Z \, \frac{m_{{l}_{i}}}{m_{{l}_j}} \,
\delta_{ij} -\frac{f(Z)}{\sqrt{2} }  \,\sqrt{\frac{m_{{l}_i}}{m_{{l}_j}}  }
\, \tilde{\chi}^{l}_{ij}  \bigg],
\label{Zij}
\end{eqnarray} }
where $G(\xi_\phi^f,X)$ and $f(x)$ can be obtained from  \cite{HernandezSanchez:2012eg} and the parameters $\xi_\phi^f$, $X$, $Y$ and $Z$ are given in the table \ref{couplings}. When the parameters $\chi_{ij}^f=0$,   one recovers the Yukawa interactions given in Refs.~\cite{Grossman:1994jb, Akeroyd:2012yg,Aoki:2009ha}.
 As it was pointed out in \cite{HernandezSanchez:2012eg}, we suggest that this Lagrangian could also represent a
Multi-Higgs Doublet Model (MHDM)
or an Aligned 2HDM (A2HDM) with additional flavor physics in the Yukawa matrices as well as the possibility of
FCNCs at tree level. Here, we present our analysis for the four versions of the 2HDM-III with a four-zero texture
introduced in the aforementioned table.
\begin{table}
\begin{center}
\begin{tabular}{|c|c|c|c|c|c|c|c|c|c|}
\hline
 2HDM-III& $X$ &  $Y$ &  $Z$ & $\xi^u_h $  & $\xi^d_h $ & $\xi^l_{h} $  & $\xi^u_H $  & $\xi^d_H $ & $\xi^{l}_H $\\ \hline
2HDM-I-like
&  $-\cot\beta$ & $\cot\beta$ & $-\cot\beta$ & $c_\alpha/s_\beta$ & $c_\alpha/s_\beta$ & $c_\alpha/s_\beta$
& $s_\alpha/s_\beta$ & $s_\alpha/s_\beta$ & $s_\alpha/s_\beta$\\
2HDM-II-like
& $\tan\beta$ & $\cot\beta$ & $\tan\beta$ & $c_\alpha/s_\beta$ & $-s_\alpha/c_\beta$ & $-s_\alpha/c_\beta$
& $s_\alpha/s_\beta$ & $c_\alpha/c_\beta$ & $c_\alpha/c_\beta$\\
2HDM-X-like
& $-\cot\beta$ & $\cot\beta$ & $\tan\beta$ &  $c_\alpha/s_\beta$ & $c_\alpha/s_\beta$ & $-s_\alpha/c_\beta$
& $s_\alpha/s_\beta$ & $s_\alpha/s_\beta$ & $c_\alpha/c_\beta$\\
2HDM-Y-like
& $\tan\beta$ & $\cot\beta$ & $-\cot\beta$ & $c_\alpha/s_\beta$ & $-s_\alpha/c_\beta$ & $c_\alpha/s_\beta$
& $s_\alpha/s_\beta$ & $c_\alpha/c_\beta$ & $s_\alpha/s_\beta$\\
\hline
\end{tabular}
\end{center}
\caption{ Parameters $\xi_\phi^f$, $X$, $Y$ and $Z$  defined in the  Yukawa interactions of eqs.   (5)--(8) for four versions of the 2HDM-III with a four-zero texture. Here $s_\alpha = \sin \alpha $, $ c_\alpha = \cos \alpha $,
$s_\beta = \sin \beta $ and $ c_\beta = \cos \beta $. }
\label{couplings}
\end{table}

\section{Feynman rules}

In this section we present the Higgs sector Lagrangian which describes
the $\gamma\gamma \phi$ and $ \gamma Z \phi$ vertices. First, we
 write the general effective Lagrangian through first order ({i.e.}, at one-loop level in perturbation theory). Then, we will
show the explicit form factors in the 2HDM-III.

The effective Lagrangian can be written as following way:
\begin{equation}\label{lagrangiano}
  {\cal L}_{\phi\gamma V}=\frac{1}{4}\Delta_{1\gamma\gamma} \phi_a
  F_{\mu\nu}F^{\mu\nu}+\frac{1}{4}\Delta_{2\gamma\gamma} A F_{\mu\nu}\widetilde
  F^{\mu\nu}+\Delta_{1\gamma Z}\phi_aF_{\mu\nu}\partial^\mu
  Z^\nu+\Delta_{2\gamma Z}A \widetilde F_{\mu\nu}\partial^\mu Z^\nu,
\end{equation}
where $\phi_a$ ($a=1,2$) is any neutral  Higgs  boson, with CP-even parity,
predicted by the model. Similarly, the $A$ represents the
neutral Higgs boson with CP-odd parity. Further, $F^{\mu\nu}$ and
$\widetilde F^{\mu\nu}$ are the electromagnetic tensor and the
dual tensor, respectively. The definitions for these tensors are:
\begin{eqnarray}
  F^{\mu\nu}&=&\partial^\mu A^\nu-\partial^\nu A^\mu,\\
  \widetilde
  F^{\mu\nu}&=&\frac{1}{2}\epsilon^{\mu\nu\alpha\beta}F_{\alpha\beta}.
\end{eqnarray}
Now, using the Lagrangian which has been presented in
eq. (\ref{lagrangiano}), the Feynman rules for $\gamma\gamma
\phi$ and $ \gamma Z \phi$ ($\phi=\phi_a,A$) can be written as:
\begin{eqnarray}
  g_{\phi_a\gamma\gamma}&=&i \Delta_{1\gamma\gamma}(k_1^\nu k_2^\mu-k_1\cdot k_2
  g^{\mu\nu})\label{vertices1},\\
  g_{A\gamma\gamma}&=&i
  \Delta_{2\gamma\gamma}\epsilon^{\mu\nu\alpha\beta}k_{1\alpha}k_{2\beta},\\
  g_{\phi_a\gamma Z}&=&i \Delta_{1\gamma Z}(k_1^\nu k_2^\mu-k_1\cdot k_2
  g^{\mu\nu}),\\
  g_{A \gamma Z}&=&i \Delta_{2\gamma
  Z}\epsilon^{\mu\nu\alpha\beta}k_{1\alpha}k_{2\beta},\label{vertices2}
\end{eqnarray}
where the $\Delta_{j \gamma V}$ (with $j=1$,2 and $V=Z$, $\gamma$) represent the form factors for the one-loop
couplings. The scheme of momentum is $k_1^\mu$ for a photon and
$k_2^\nu$ for the second photon or the $Z$ boson.
Finally, the tensor amplitudes are obtained from eqs.
(\ref{vertices1})--(\ref{vertices2}) and these can be written in terms of the
CP-even and CP-odd parts,
\begin{eqnarray}
  {\cal M}^{\mu\nu}_{even}&=&i\Delta_{1\gamma V}(k_1^\nu k_2^\mu-k_1\cdot k_2
  g^{\mu\nu}),\\
  {\cal M}^{\mu\nu}_{odd}&=&i\Delta_{2\gamma
  V}\epsilon^{\mu\nu\alpha\beta}k_{1\alpha}k_{2\beta},
\end{eqnarray}
Now, the decay $\Gamma(\phi_i\to\gamma V)$ can be completely determined considering the explicit forms of
$\Delta_{i\gamma V}$ for the 2HDM-III which are presented in the two upcoming subsections
 where we have introduced the following  notation:  $V= \gamma, Z$ and $i=1,2$ with $i=1$ for $\phi_a$
(which in turn refers to either $h$ or $H$)
 and $i=2$ for $A$. The explicit expressions for the two decays studied are shown in Appendix A.

\subsection{Form Factor $\Delta_{1\gamma\gamma}$}

\begin{figure}[ht]
\centering
\includegraphics[width=5 in]{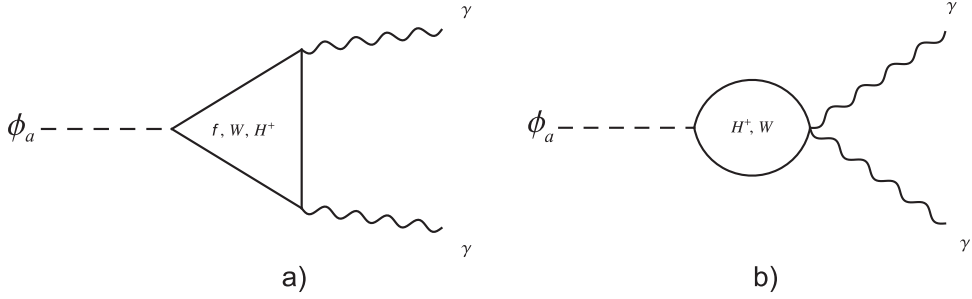}
\caption{The Feynman diagrams for the $\phi_a\gamma\gamma$ vertex.}\label{phiagg}
\end{figure}

Here, we present the explicit expressions for the
$\Delta_{i\gamma\gamma}$ form factor in the 2HDM-III. This form factor
receives contributions from all charged particles, for this reason it
is convenient to separate each sector:
\begin{equation}
\Delta_{1\gamma\gamma}=\Delta^0_{1\gamma\gamma}+\Delta^1_{1\gamma\gamma}+\Delta^{1/2}_{1\gamma\gamma}.
\end{equation}
In the last equation we have labelled with $0$ the contribution from the
scalar sector, with $1/2$ from the fermionic sector and with $1$ from the gauge
sector. In an explicit way, these contributions are (refer to figure ~\ref{phiagg}):
\begin{eqnarray}
\Delta^0_{1\gamma\gamma}&=&\frac{-\alpha^{3/2} m_W }{4\pi^{1/2}s_Wk_1\cdot k_2}\Big[2m_{H^\pm}^2C_0(1,2)+1\Big]{\cal G}_{\phi_i H^\pm H^\mp},
\\
\Delta^1_{1\gamma\gamma}&=&\frac{-\alpha^{3/2}}{\pi^{1/2}m_W s_W k_1\cdot k_2}
\Big[6 m_W^2(m_W^2-k_1\cdot k_2)C_0(1,2)\nonumber\\
&&+k_1\cdot k_2+3m_W^2\Big]{\cal G}_{\phi_iWW},\\
\Delta^{1/2}_{1\gamma\gamma}&=&\sum_f\frac{2\alpha^{3/2}N_c m_f^2
Q_f^2}{\pi^{1/2}m_Ws_W k_1\cdot k_2}\Big[(2m_f^2-k_1\cdot
k_2)C_0(1,2)+1\Big]{\cal G}_{\phi_a\bar f f}.
\end{eqnarray}
Here, we have introduced two shorthand notations. The first is
${\cal G}_{ABC}$, which represents the dimensionless function related to
the couplings between the particles $ABC$
(see Appendix B). The second shorthand notation is for the
Passarino-Veltman functions, that is
\begin{equation}
C_0(a,b)=C_0(k_a^2,k_b^2,2k_a\cdot k_b,m^2,m^2,m^2),
\end{equation}
where $m^2$ has to be taken according to every particle in the
loop.

\subsection{Form Factor $\Delta_{2\gamma\gamma}$}

\begin{figure}
\centering
\includegraphics[width=2in]{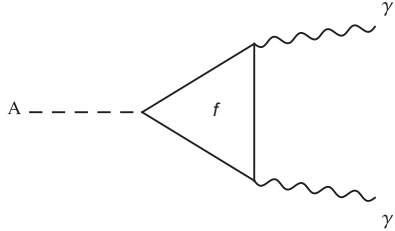}\\
\caption{The Feynman diagram for the $A\gamma\gamma$ vertex.}\label{Agg1}
\end{figure}

This form factor, due to the presence of a Higgs boson $A$, only
receives contributions from the fermionic sector. These contributions
are introduced through the Feynman diagram shown in figure
\ref{Agg1}. The explicit expression for this form factor is:
\begin{equation}
\Delta_{2\gamma\gamma}=\sum_f\frac{-4ie^3 N_c m_f^2 Q_f^2}{m_W
s_W}C_0(1,2){\cal G}_{A\bar f f}.
\end{equation}

\subsection{Form Factor $\Delta_{1\gamma Z}$}

\begin{figure}[ht]
\centering
  \includegraphics[width=5in]{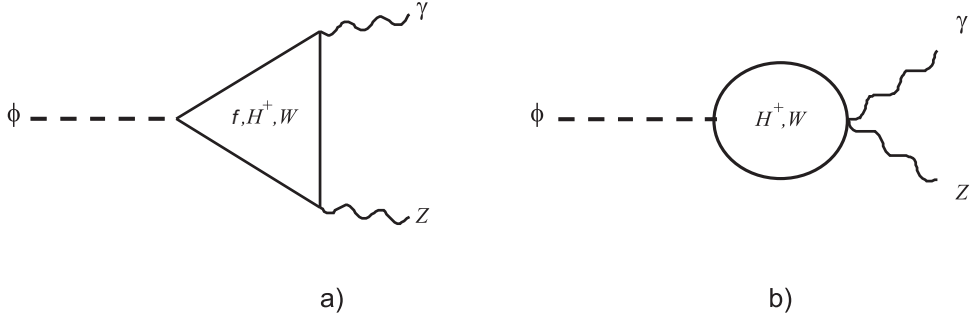}\\
  \caption{The Feynman diagrams for the $\phi Z\gamma$ vertex.}\label{phiZg1}
\end{figure}

Now, for the $\gamma Z \phi_a$ vertex, we will have again contributions
from all charged particles, see figure \ref{phiZg1}. Therefore,  it is again convenient to
separate every contribution. Hence, we have
\begin{eqnarray}
\Delta_{1\gamma Z}^0&=&\frac{-m_W \alpha^{3/2}c_{2W}}{8\pi^{1/2}s_Ws_{2W}(k_1\cdot k_2)^2}
\left\{k_1\cdot k_2\Big[4m_{H^\pm}^2C_0(1,2)+2\Big]\right.\nonumber\\
&&\left.+m_Z^2\Big[B_0(P)-B_0(k_2)\Big]\right\}{\cal G}_{\phi H^\pm H^\mp},\\
\Delta_{1\gamma Z}^1&=&\frac{-c_W \alpha^{3/2}}{8\pi^{1/2}m_W^3 s_W^2 (k_1\cdot k_2)^2} {\cal G}_{\phi_a WW}\left\{4C_0(1,2)m_W^2
k_1\cdot k_2\Big[2(k_2^2-6m_W^2)k_1\cdot k_2\right.\nonumber\\
&&-k_2^2+12m_W^4\Big]-\Big[(2k_2^2-4m_W^2)k_1\cdot k_2+k_2^2-12 m_W^4\Big]\nonumber\\
&&\left.\times(k_2^2[B_0(P)-B_0(k_2)]+2k_1\cdot k_2)\right\},
\\
\Delta_{1\gamma Z}^{1/2}&=&\sum_f\frac{f_V N_c m_f^2 Q_f \alpha^{3/2}}{4\pi^{1/2}c_Ws_W^2m_W(k_1\cdot k_2)^2}
\left\{2k_1\cdot k_2\Big[(2m_f^2-k_1\cdot k_2)C_0(1,2)+1\Big]\right.\nonumber\\
&&\left.+m_Z^2\Big[B_0(P)-B_0(k_2)\Big]\right\}{\cal G}_{\phi ff}.
\end{eqnarray}
Here, $f_V$ is the vector part of the coupling $\bar f f Z$ (see its
explicit form in Appendix B). Also, we have used a shorthand notation
for the $B_0$ Passarino-Veltman function, this is
\begin{equation}
B_0(k)=B_0(k \cdot k,m^2,m^2).
\end{equation}

\subsection{Form Factor $\Delta_{2\gamma Z}$}

\begin{figure}[ht]
\centering
  \includegraphics[width=2in]{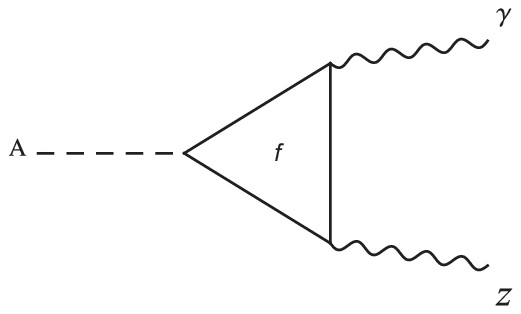}\\
  \caption{The Feynman diagram for the $A Z\gamma$ vertex.}\label{AZg1}
\end{figure}

For this case, similarly to $\Delta_{2\gamma\gamma}$, the
form factor receives contributions only from the fermionic sector, see
figure  \ref{AZg1}. Explicitly, we can write as follows:
\begin{equation}
\Delta_{2\gamma Z}=\sum_f\frac{-ie^3 f_V m_f^2
Q_fN_c}{c_Ws_W^2m_W}C(1,2){\cal G}_{A\bar f f}.
\end{equation}

\section{Discussion}

In this section we will present the results for the  two Brs of interest, {i.e.}, of the channels
$\phi \to \gamma \gamma$ and $\gamma Z$, where (again) $\phi=\phi_a$ or $A$. Recently,
some of us have studied the flavor constraints affecting the 2HDM-III and we have
isolated the surviving parameter space \cite{HernandezSanchez:2012eg} (again, see also \cite{Felix-Beltran:2013tra}), which we are going to re-use in our present analysis.    Besides,  we will
incorporate an extensive discussion of the most popular models, like 2HDM-I, 2HDM-II, 2HDM-X and 2HDM-Y, which
are particular incarnations of our 2HDM-III.
However, do recall that experimental results suggest a SM-like Higgs signal, for this reason we have therefore chosen the following scenario
\begin{eqnarray}\label{alpha-beta}
\beta-\alpha&=&\frac{\pi}{2}+\delta,\\
\lambda_6 &=& -\lambda_7 \\
\mu_{12}&\sim& v,
\end{eqnarray}
where $\delta$ is considered near to zero and where we take $\mu_{12}=200$ GeV. Besides, we can observe that is more convenient to use $\lambda_6= - \lambda_7$ instead of $\lambda_6=  \lambda_7$ because the rates of $h \to \gamma \gamma$, $\gamma Z$ can receive  the greatest enhancement. In the opposite case, $\lambda_6=  \lambda_7$,
  the contribution to the decay is irrelevant (see the three Higgs bosons vertices Feynman rules of appendix B).  So that our settings
naturally comply with the SM-like scenario advocated in Ref.~\cite{Gunion:2002zf}.

\subsection{The $h \to \gamma \gamma, \, \gamma Z$ decays}

\begin{figure}[ht]
  \centering
  \includegraphics[width=2.8in]{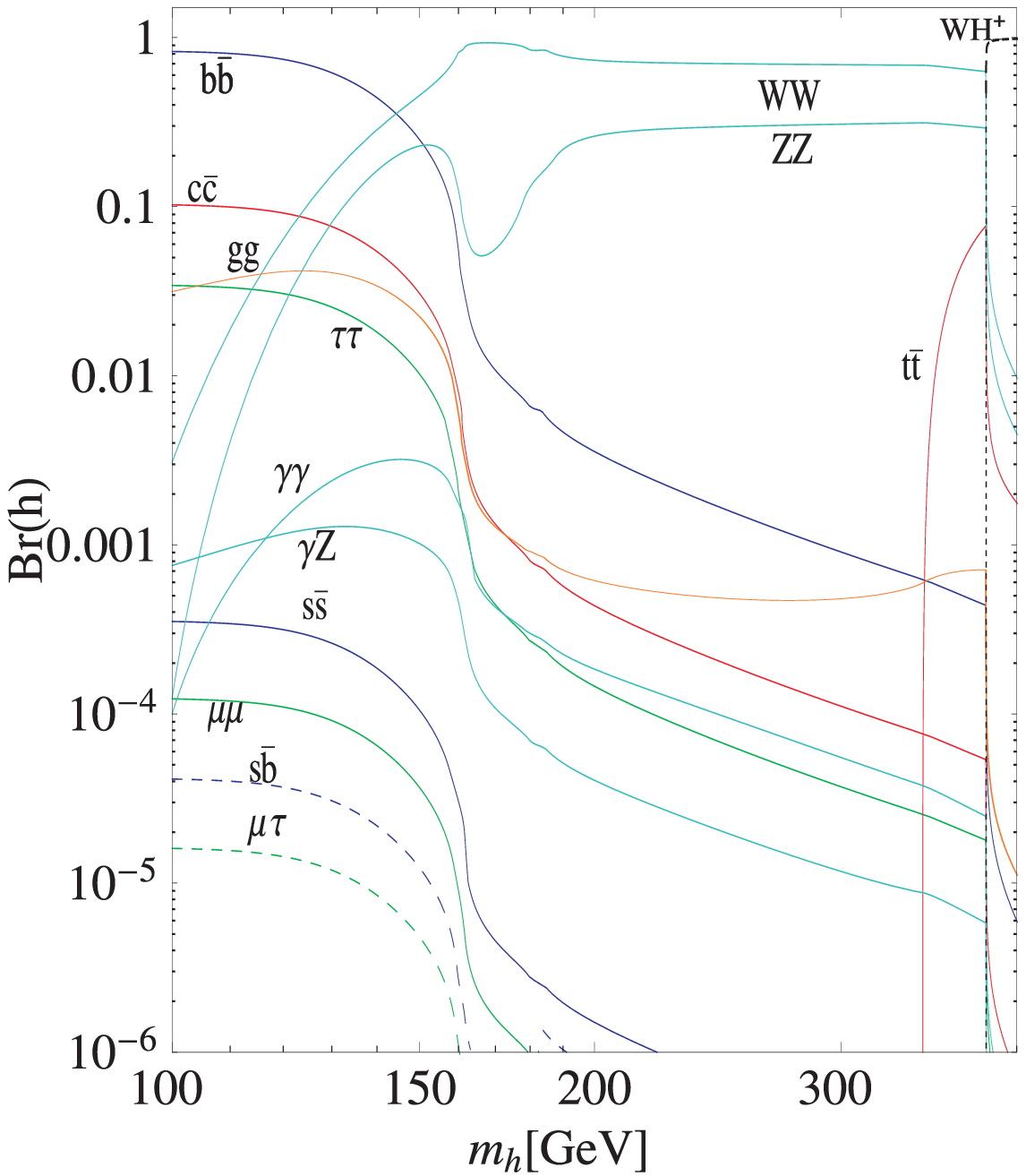}
  \includegraphics[width=2.8in]{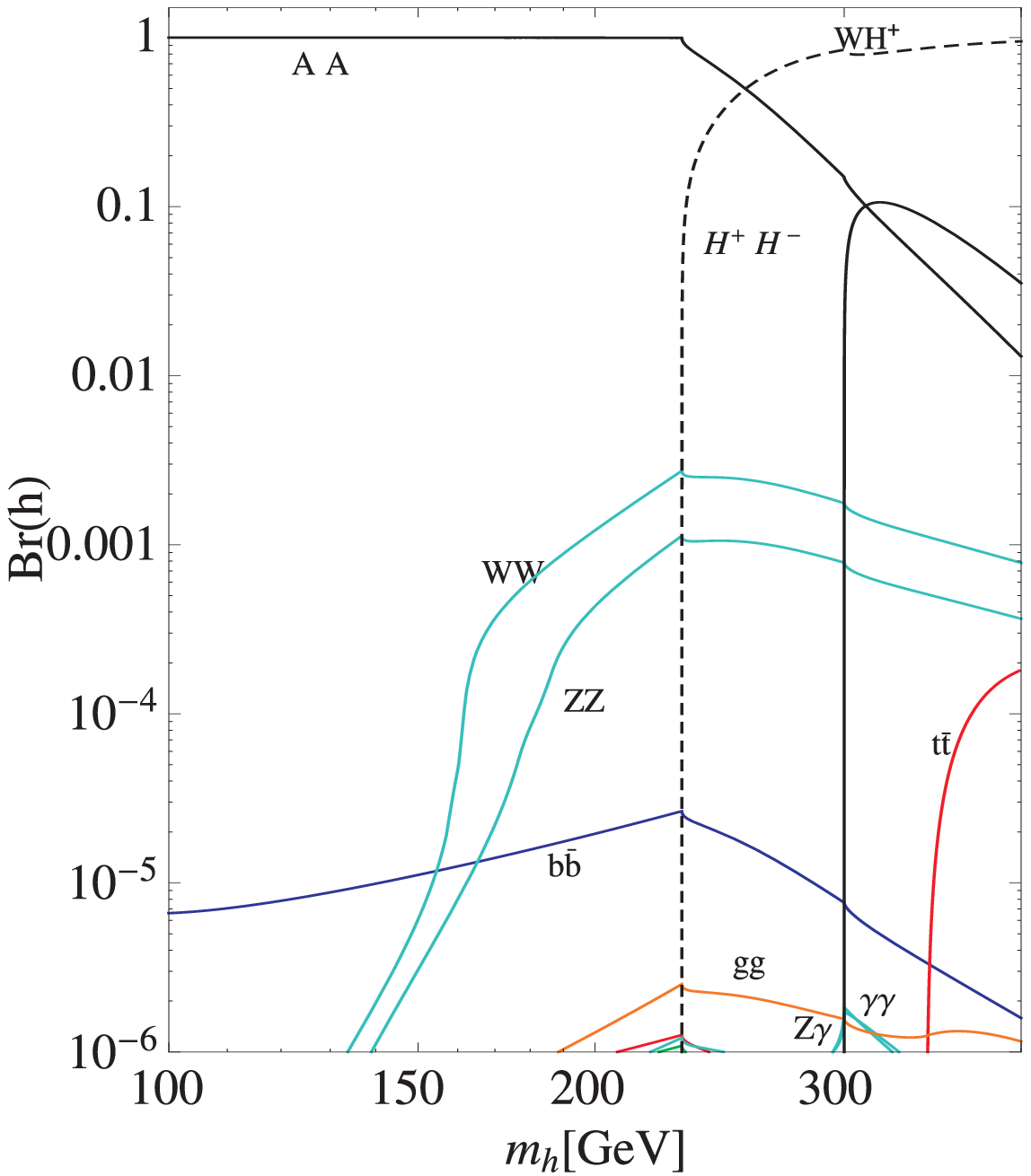}\\
  \caption{Behavior of all decay channels of the light CP-even Higgs boson $h$ with respect to its mass for the
 2HDM-III Like II. The parameters used are as
follows:   $\chi_{kk}^u=\chi_{kk}^d=1$,  $\chi_{23}^u=-0.75$, $\lambda_7=-\lambda_6=-1$, $X=10$ for  (a)  $m_A>m_h$, $m_{H^+}=300$ GeV and $\chi_{23}^d=-0.035$ (left panel) and  (b) $m_A=40$ GeV, $m_{H^+}=150$ GeV and  $\chi_{23}^d=0.002$ (right panel).
}\label{allbran1}
\end{figure}

In this section we present the results for the case of $h$ decays. We start with a general discussion of all decay channels and we finally comment on the  two specific channels of interest.
In the left panel  of  figure \ref{allbran1}, where the $h\to AA$ decay is forbidden,   one can see that the behavior of all decay channels is similar to the SM case \cite{Djouadi:2005gi}. However, if the decay $h\to AA$ is kinetically allowed (see right panel), all SM channels show a strong reduction, as this mode becomes dominant for most $m_h$ values.  For this special case ($m_A < m_h/2$ ), there is a small region of parameter space of our model, where this channel decay is allowed. Following the study of new physics effects on the electroweak oblique parameters parametrized by S, T and U \cite{Kanemura:2011sj}, we find  for  $2 m_A < m_h $ and $m_H \sim 200$ - 230 GeV, taking $\sin (\beta - \alpha) \sim 1$, the  range allowed for the charged Higgs boson mass is given by 150 GeV$ \leq m_{H^\pm} \leq 200$ GeV. Using these values for the masses of neutral and charged  Higgs bosons, we can confront the parameter space of our model  with the main flavor physics constraints, which are studied in \cite{HernandezSanchez:2012eg,Crivellin:2013wna}. We can obtain practically the same constraints for the parameters of Yukawa matrices with a four-zero texture, except for the off-diagonal term, $\chi_{23}^d$, which must be very tiny  and it has the following bound
$| \chi_{23}^d | \leq 10^{-1} $. The process $B_s \to \mu^+ \mu^-$ imposes the most strong  constraint to the parameter $\chi_{23}^d$ (see the formula of this process in the Ref. \cite{HernandezSanchez:2012eg}).   On the other hand, we should consider another assumption, the  possibility to observe this channel decay at LHC.  In Ref. \cite{Carena:2007jk} the decay $h \to AA$ is studied in a model-independent way with $2 m_A < (m_h -10)$ GeV, this channel could provide  sizable significances for an integrated luminosity $L = 30 \, fb^{-1}$ and adequate b-tagging efficiencies.
Therefore, if we want to have a $h$ boson that be SM-like, we have to demand that $2m_A>m_h$, so that the decay $h\to AA$ is forbidden.
For reference, hereafter, we are using the
 2HDM-III Like II (for reasons which will become clear below).

\begin{figure}[ht]
  \centering
  \includegraphics[width=5in]{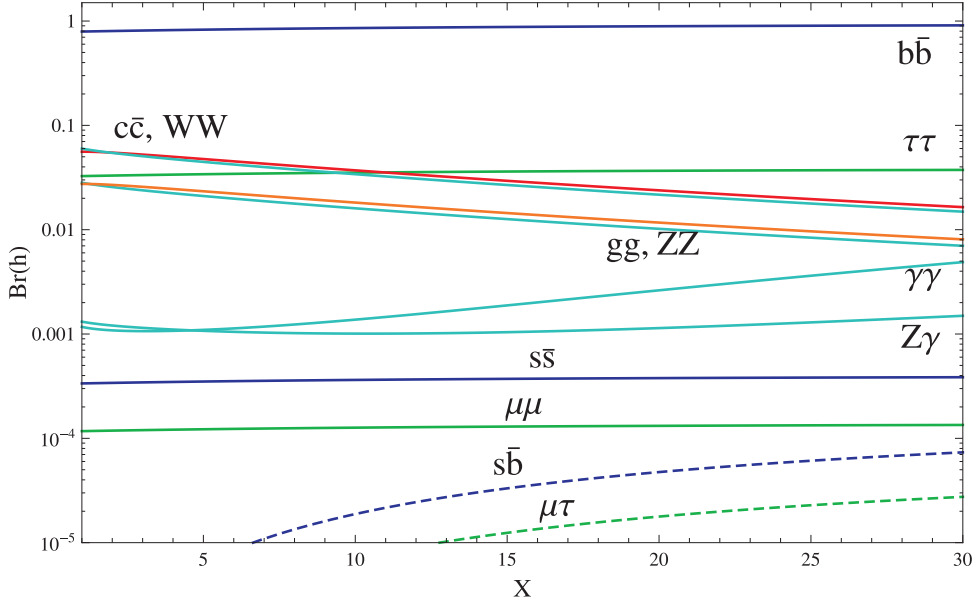}\\
  \caption{Behavior of all decay channels of the light CP-even Higgs boson $h$ with respect to the $X$ parameter of eq.
(\ref{Xij}) for the
 2HDM-III Like II. The other parameters  are the same as in the left frame of figure \ref{allbran1}.
}\label{allbran2}
\end{figure}

As we can see in figure \ref{allbran2}, the Br$(h\to\gamma\gamma)$ is very sensitive to the $X$ parameter given in eq. (\ref{Xij}). For large values of the latter, in particular, the Br$(h\to\gamma\gamma)$  shows an enhancement of one order of magnitude, but this behavior is contrary to the experimental results from the LHC. In contrast, for medium values of $X$ (say, $X<15$), this increase is under control, indeed compatible with the LHC data, so that  we will choose a definite
value in this range, e.g., $X=10$, from now on. We will instead change the values of other parameters, like the mass of the
charged Higgs boson, $m_{H^+}$.

In the remainder of this subsection, we  analyze $h\to \gamma\gamma$ and $\gamma Z$ relative to the SM
case, by introducing the so-called  $R$ parameters,
\begin{eqnarray}
  R_{\gamma X}&=&\frac{\sigma(gg\to h)|_{\rm 2HDM-III}\times {\rm Br}(h\to \gamma X)|_{\rm 2HDM-III}}{\sigma(gg\to h)|_{\rm SM}\times {\rm Br}(h\to \gamma X)|_{\rm SM}}\nonumber\\
  &\approx&{\cal G}_{htt}^2\frac{{\rm Br}(h\to \gamma X)|_{\rm 2HDM-III}}{{\rm Br}(h\to \gamma X)|_{\rm SM}} \, \, \, (X= \gamma, Z),
\end{eqnarray}
where $ {\cal G} _ {htt} $ is the ratio of the couplings $ htt | _ {\rm 2HDM-III} $ and $ htt | _ {\rm SM} $
entering the $hgg$ effective vertex (see Appendix A).
Notice that, in the case of a fermiophobic $h$ state, the $gg\to h$ production mode ought to be replaced by either
vector boson fusion or Higgs-strahlung, for which the ratio of cross sections reduces to unity, so that the above
formula remains applicable upon the replacement  ${\cal G}_{htt}\to 1$.

\begin{figure}[ht]
  \centering
  \includegraphics[width=3in]{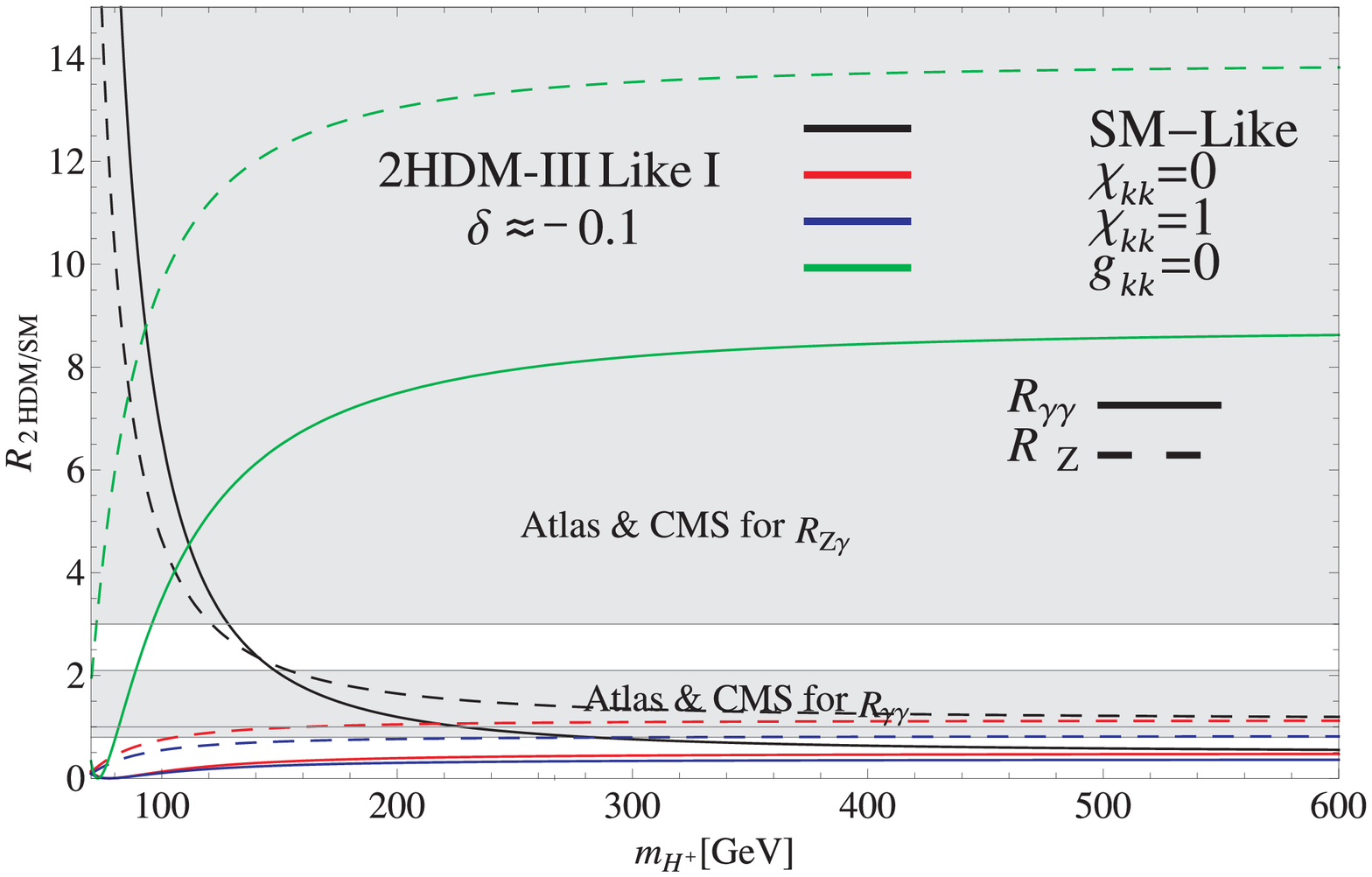}\includegraphics[width=3in]{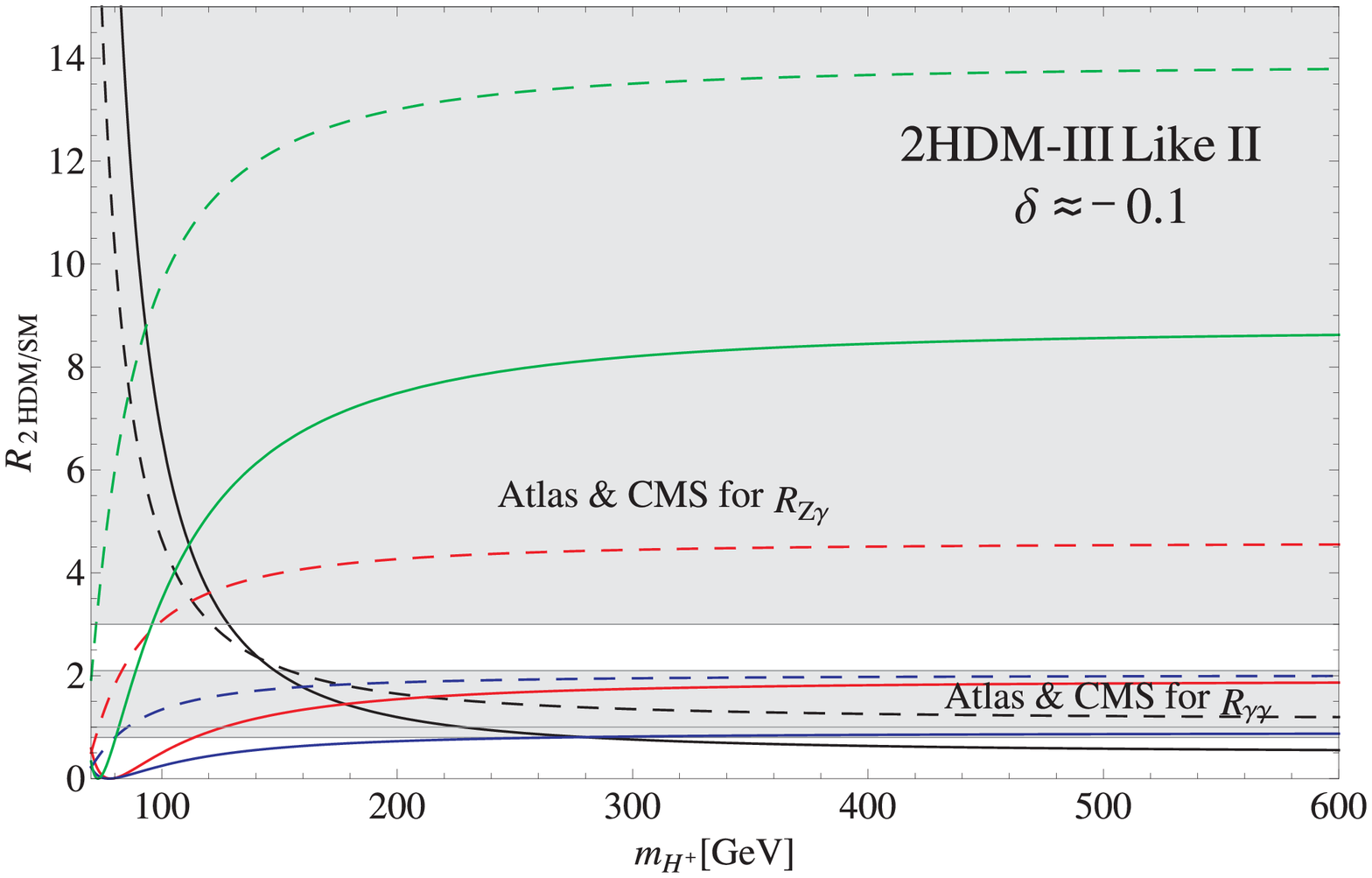}\\
  \includegraphics[width=3in]{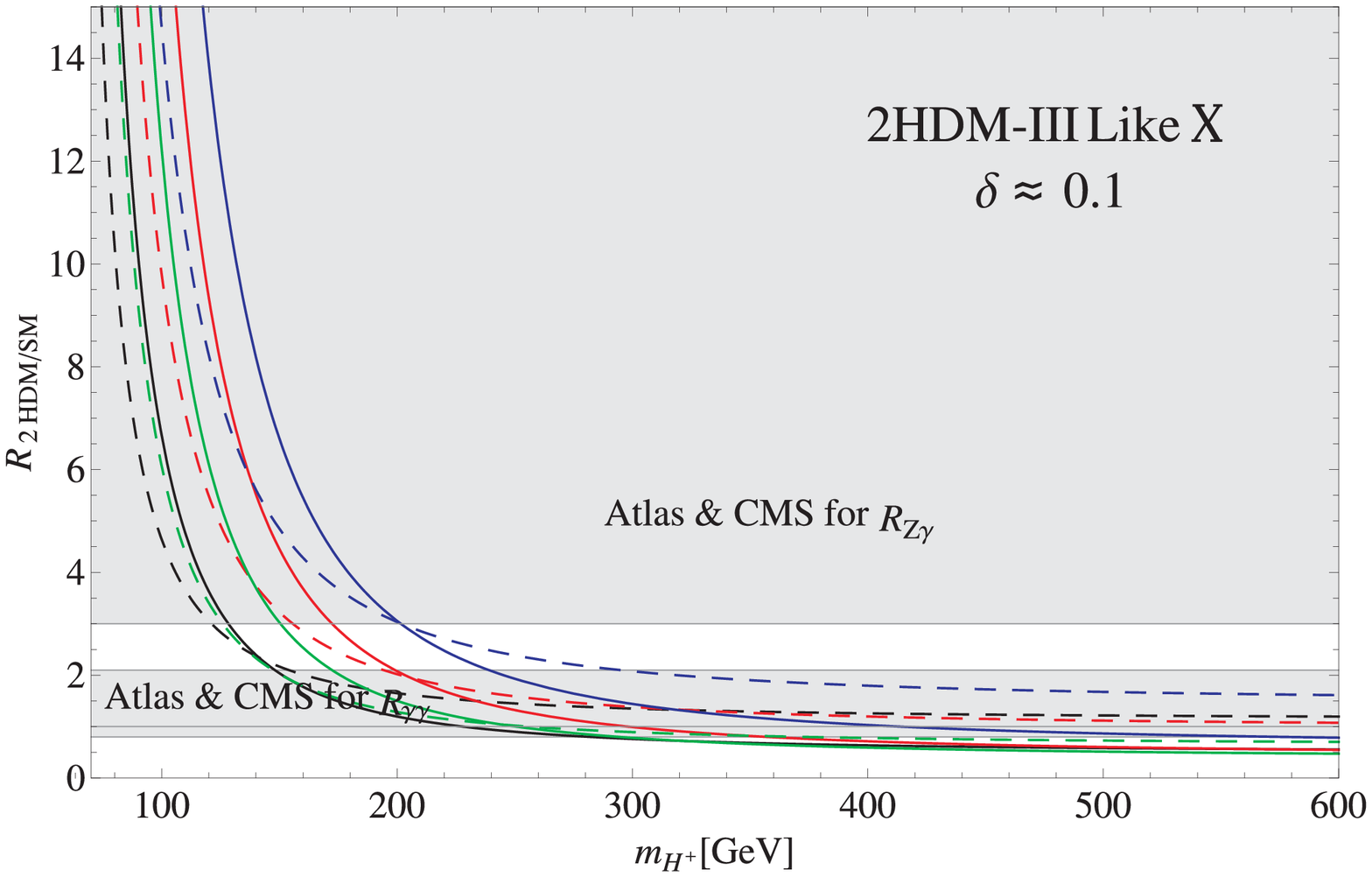}\includegraphics[width=3in]{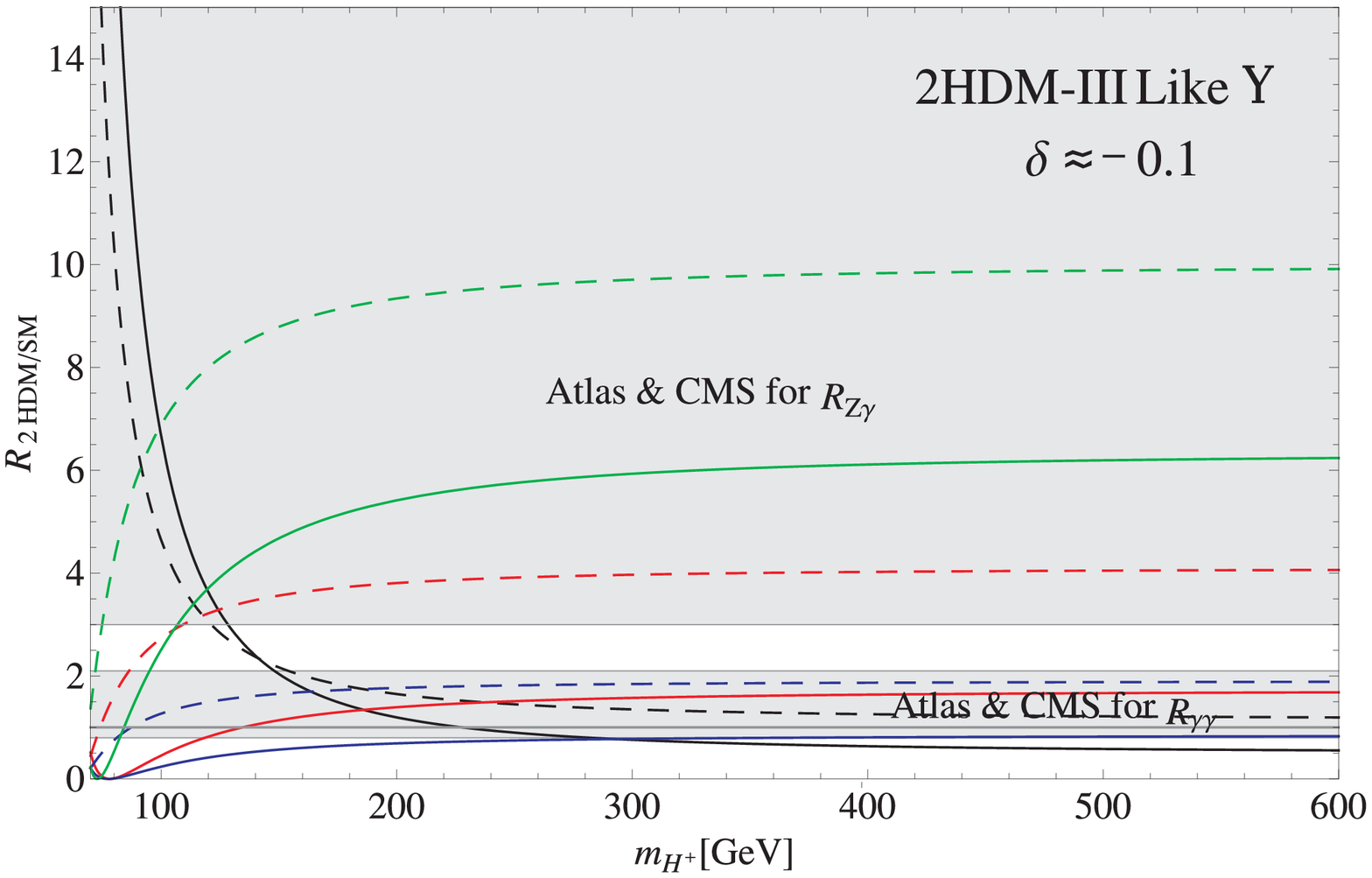}
  \caption{$R_{\gamma\gamma}$ (solid-line) and $R_{\gamma Z}$ (dashes-line) with respect to the charged Higgs boson mass.   In all cases 
  $\lambda_7= -\lambda_6=-1$.  The other parameters  are given in the legends and described in the text. }\label{Rh}
\end{figure}

In figure \ref{Rh} we show the behavior  of $R_{\gamma\gamma}$ and $R_{\gamma Z}$ with respect to the charged  Higgs boson mass, $m_{H^+}$. In the plots,  the shaded areas represent the fits to the experimental results
from the LHC. In particular, the scenarios presented are the following: the black line is for an exactly SM-like
$h$ state ($\delta=0$), the red line represents the case when the Yukawa couplings are equal to the 2HDM with $\mathcal{Z}_2$ symmetry, the blue line is associated to a set of Yukawa couplings with FCNCs ($\chi_{23}^d=-0.35$ and $\chi_{23}^u=-0.75$), finally, the green line illustrates the fermiophobic  scenario.

One can see in the figure that the most relevant scenarios are:  2HDM-III-like II and Y  with $\chi_{kk}^f=0$ and the fermiophobic scenario for 2HDM-III-like I, II and Y. The parameterisations 2HDM-III-like X is disadvantaged for all scenarios presented. The fermiophobic  scenario demands a charged Higgs boson very light, between $80$ and $90$ GeV for the 2HDM-like I, II and Y. Notice that the $\chi_{kk}^f=0$ scenario opens up the possibility of a light charged Higgs boson, $m_{H^+} \geq 110$ GeV, for 2HDM-III-like II and Y, as already seen in \cite{HernandezSanchez:2012eg}.

\begin{figure}[ht]
  \centering
  \includegraphics[width=2.9in]{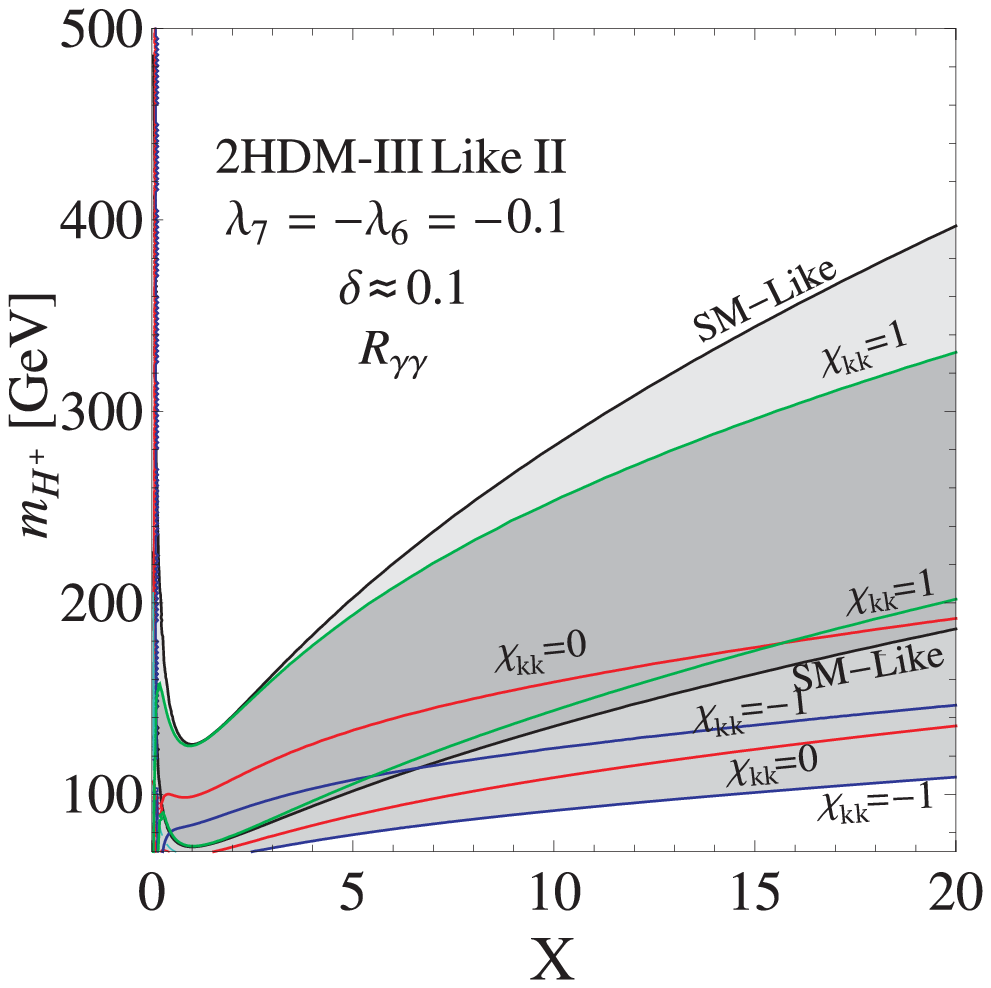}\includegraphics[width=2.9in]{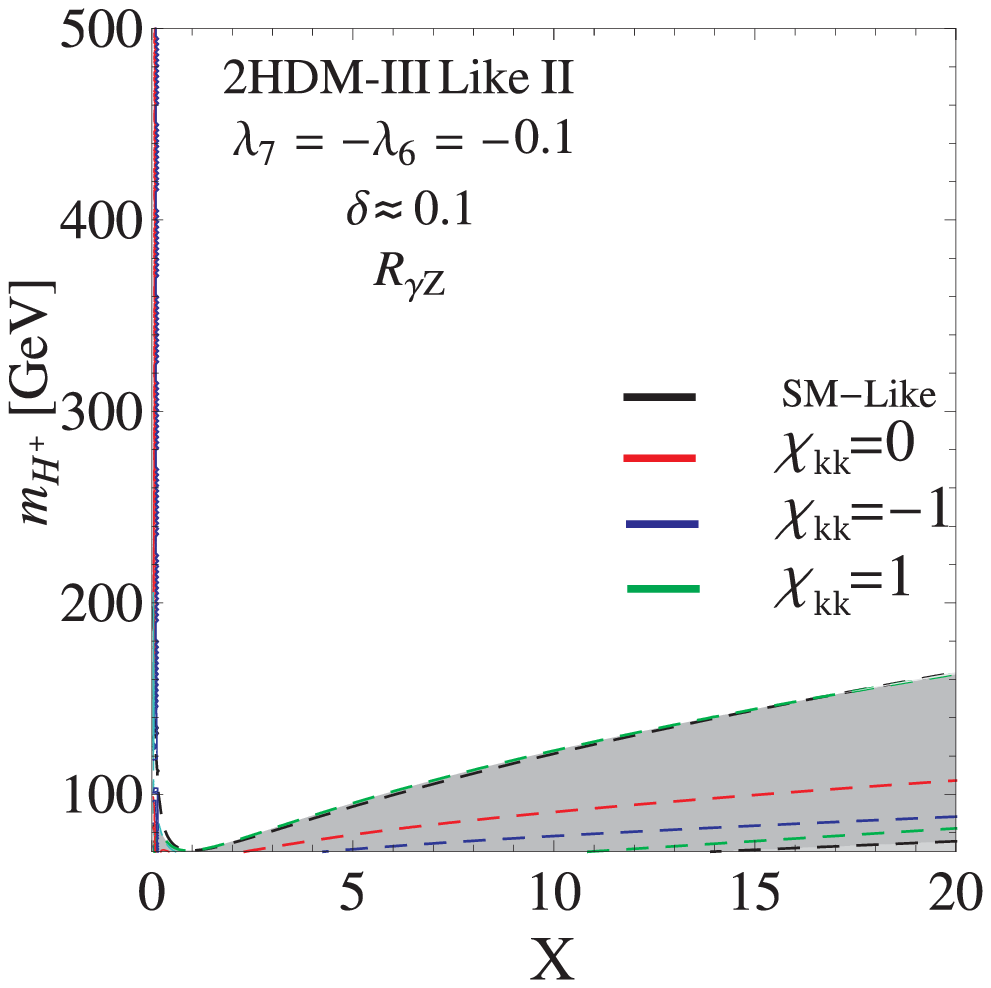}\\
  \includegraphics[width=2.9in]{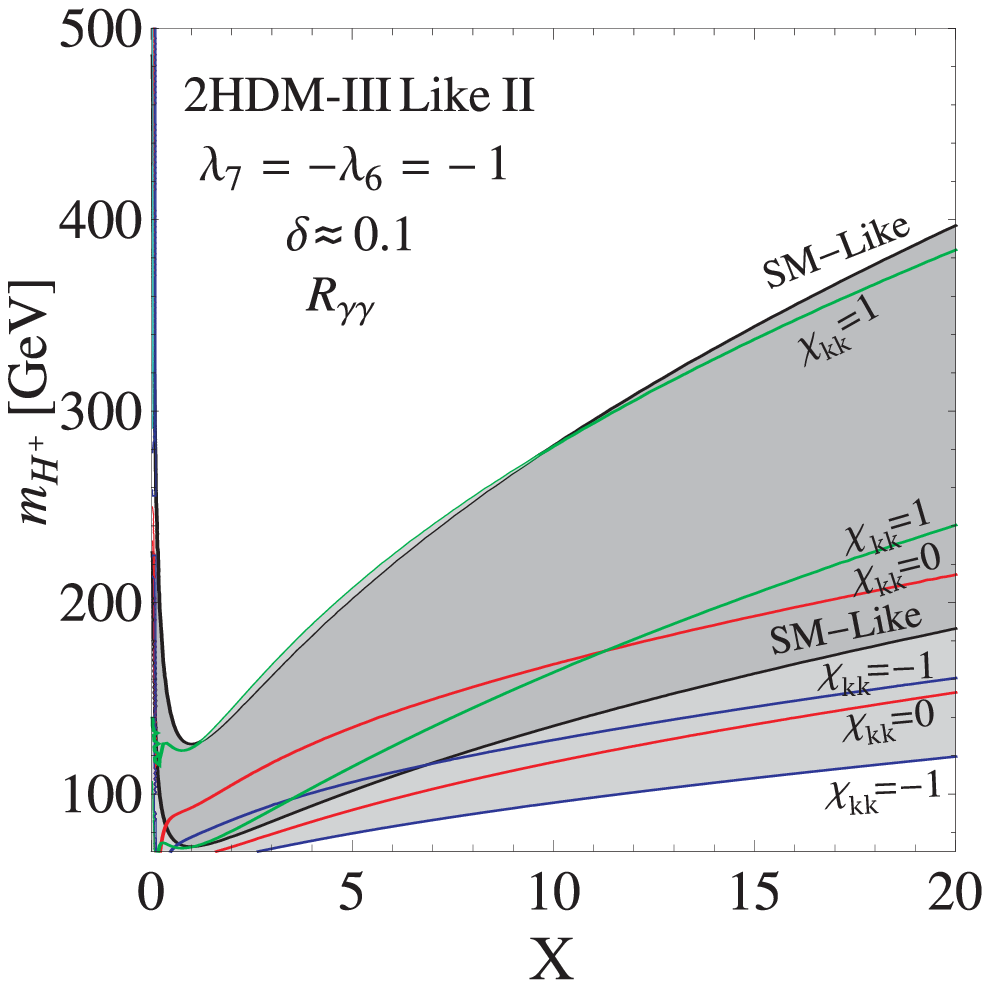}\includegraphics[width=2.9in]{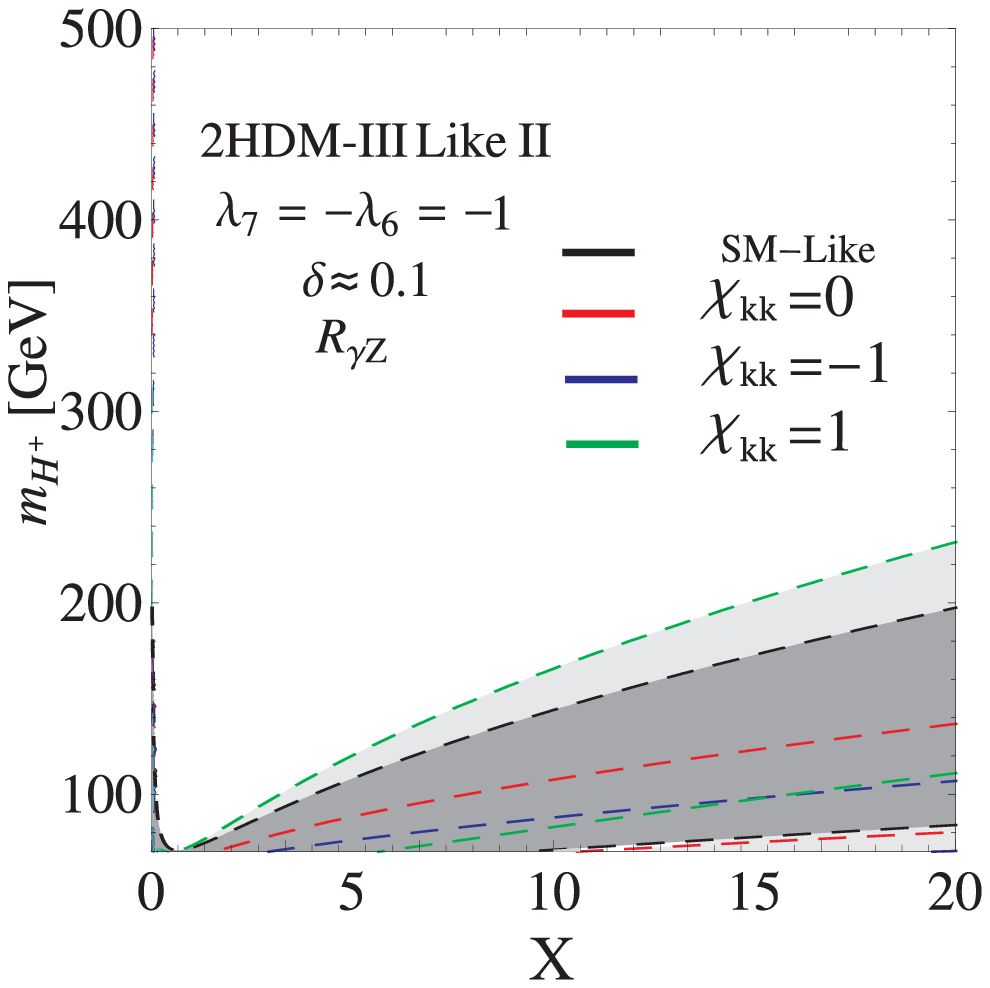}
  \caption{Constraints over the ($m_{H^+}$, $X\equiv\tan\beta$) parameter space  of the 2HDM-III-like II derived from $R_{\gamma\gamma}$ (solid-line) and $R_{\gamma Z}$ (dashes-line) measurements at the LHC ( the shaded areas and enclosed by lines of the same color, are the allowed permitted region by CMS and ATLAS). The values of $\lambda_{6,7}$ and $\delta$  are given in the legends.}\label{mc-x}
\end{figure}

Because the most important signatures are generated into the context of the 2HDM-III-like II and Y, from now on we are going to
systematically focus on 2HDM-III-like II. Within this scenario,  we present in figure \ref{mc-x} the allowed
 parameter space (after enforcing the LHC constraints) mapped onto the $m_{H^+}-X$ plane, for two values of $\lambda_{{6,7}}$
and a definite choice of $\delta$ (here, $X=\tan\beta$: see table \ref{couplings}).
As we can see, the final state $\gamma Z$ is the most constrained one, in the sense that LHC results are reproduced in a
smaller region of parameters with respect to the case of $\gamma\gamma$. In particular, the former decay demands a mass of the charged Higgs boson below
$\approx 160$  GeV for $X=20$ and $\lambda_7=-\lambda_6=-0.1$ and below  $230$ GeV for $\lambda_7=-\lambda_6=-1$. This is a valuable result, as charged Higgs bosons with such a mass and with
$X=20$ are indeed accessible at the LHC (albeit at high energy and luminosity only). In fact,
for more acceptable values of $X$ which are mid-range, e.g. around 15, we find that $m_{H^+}< m_t$, so that this state is copiously produced in top quark decays. Again, as already emphasized in Ref. \cite{HernandezSanchez:2012eg},
a light charged Higgs boson could be a hallmark manifestation of a 2HDM-III.

\begin{figure}[ht]
  \centering
  \includegraphics[width=2.9in]{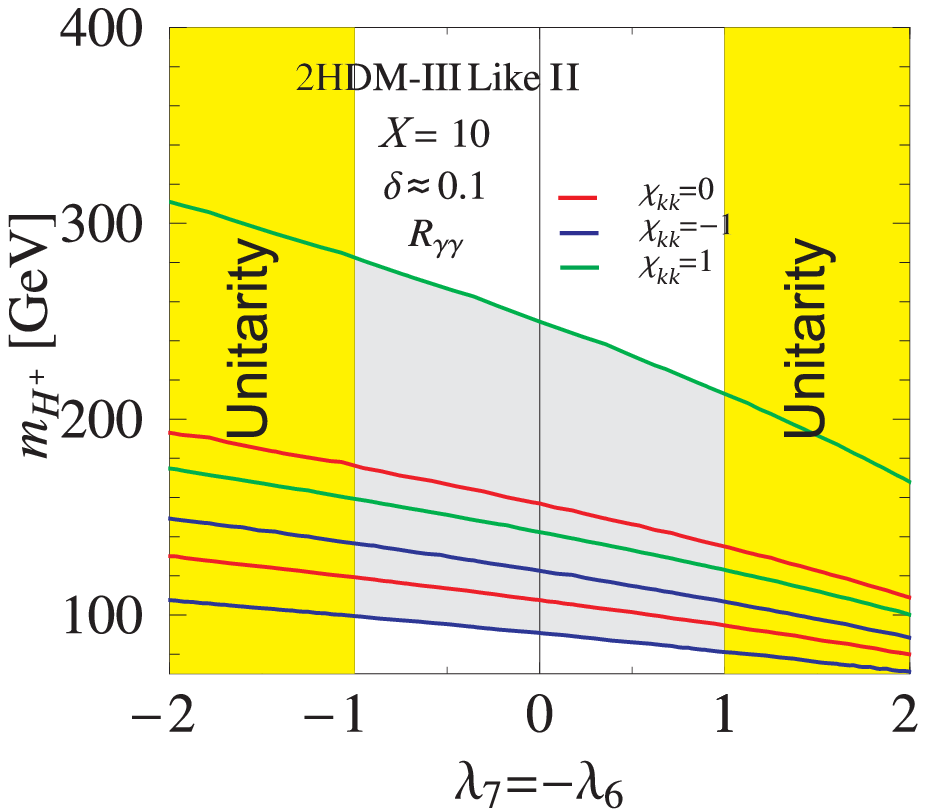} \includegraphics[width=2.9in]{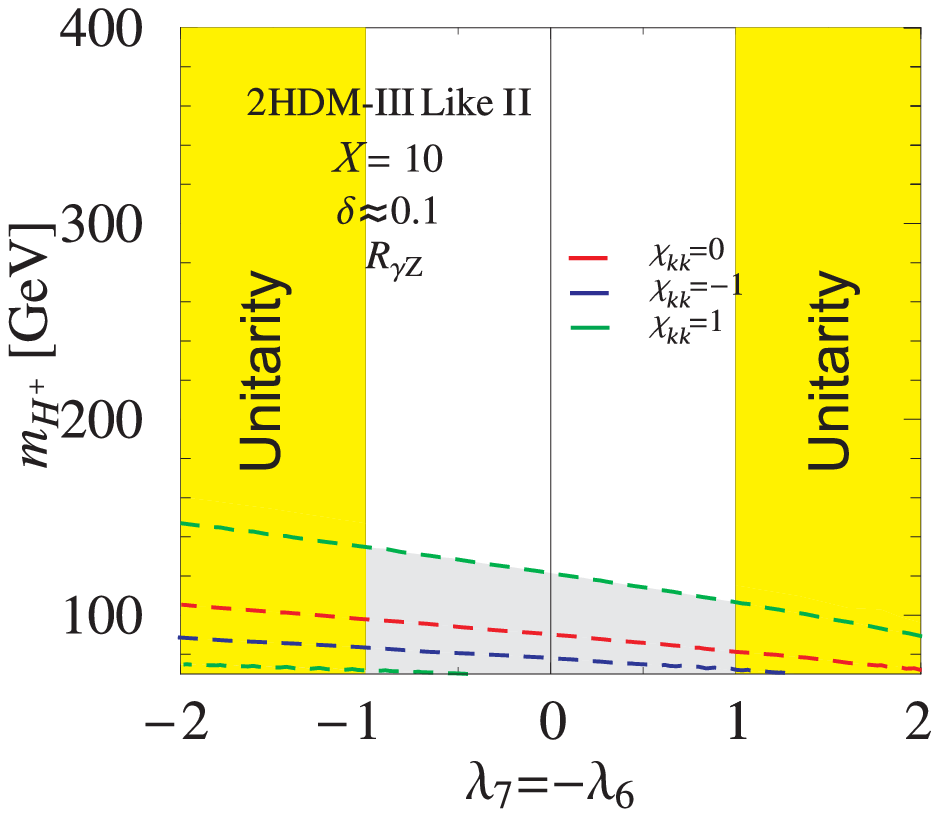}\\
  \caption{Constraints over the ($m_{H^+}$, $\lambda_7=-\lambda_6$) parameter space of the 2HDM-III-like II
derived from $R_{\gamma\gamma}$ (solid-line) and $R_{\gamma Z}$ (dashes-line). Again, 
 the shaded areas and enclosed by lines of the same color, are the allowed region by LHC. The yellow region is not allowed by constraints of tree-level unitarity}\label{mc-l}
\end{figure}

Finally, in the graphics presented in figure \ref{mc-l}, we map the same parameter space described by the previous plot now
in terms of the plane ($m_{H^+}$ , $\lambda_7=-\lambda_6$).  As we can see, when $\lambda_{{6,7}}=0$, the overlapping areas required by the decays
 $ h \to \gamma \gamma$ and $ \gamma Z $ suggest a $H^+$ mass around $100-150$ GeV, however, for $\lambda_7=-\lambda_6=-1$, the limit for
 this mass goes up to $160$ GeV, in line with our previous findings. Besides, we show in the yellow area  region excluded  by tree-level unitarity. 

\subsection{The $H \to \gamma \gamma, \, \gamma Z $ decays}

In this subsection, we present the results for the Brs of the  heavy CP-even Higgs state, denoted by $H$. We present them only
for the case of the 2HDM-III Like II, because this scenario allows for the existence of a light
charged  Higgs boson ($m_{H^+}\sim100 $ GeV), a key signature of this scenario which will be accessible at the LHC, as previously explained.

\begin{figure}[ht]
  \centering
  \includegraphics[width=5in]{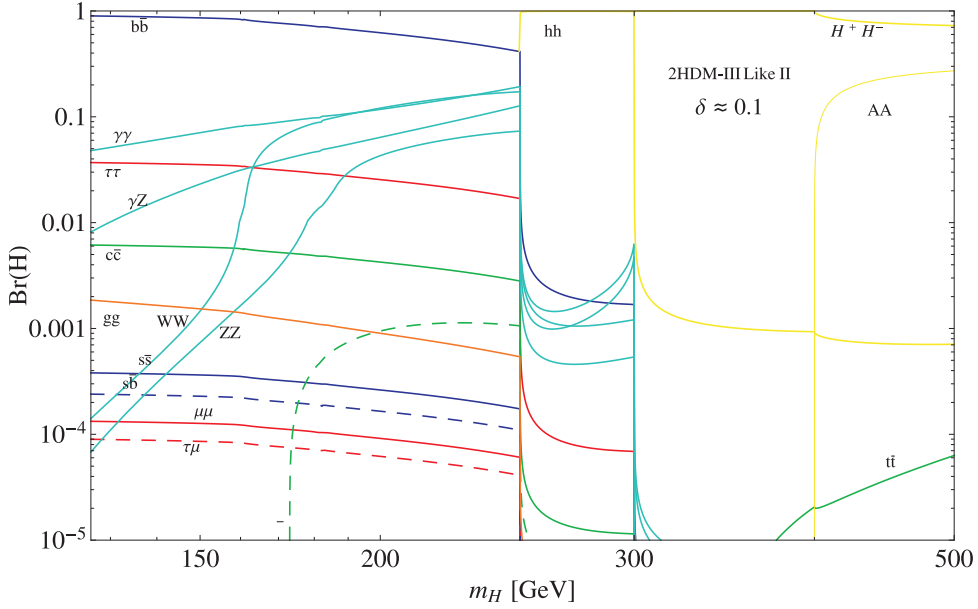}\\
  \caption{Behavior of all decay channels of the heavy CP-even Higgs boson $H$ with respect to its mass
for the
 2HDM-III Like II. The parameters used are as
follows: $m_A=200$ GeV, $m_{H^+}=150$ GeV, $\lambda_7=-\lambda_6=-1$, $\chi_{kk}^f=1$, $\chi_{23}^f=-0.35$,
 $\chi_{23}^u=-0.75$ and $X=10$.
 }\label{all-Hp}
\end{figure}

In figure \ref{all-Hp} we present all the decay channels of the heavy Higgs state. Herein, as we can see, Higgs-to-Higgs decays can again be dominant, whenever $m_H>2m_h,2m_A, 2m_{H^+}$, as the  channels $H\to hh, AA, H^+H^-$
overwhelm all others. However, in the $m_H$ region where these channels are forbidden, the final states
$\gamma\gamma$ and $\gamma Z$ turn out to be very important, becoming order of $10^{-1}$,
a significant increase above and beyond the SM rates, and only second in size to the $H\to b\bar b$ mode.

\begin{figure}
  \centering
  \includegraphics[width=2.9in]{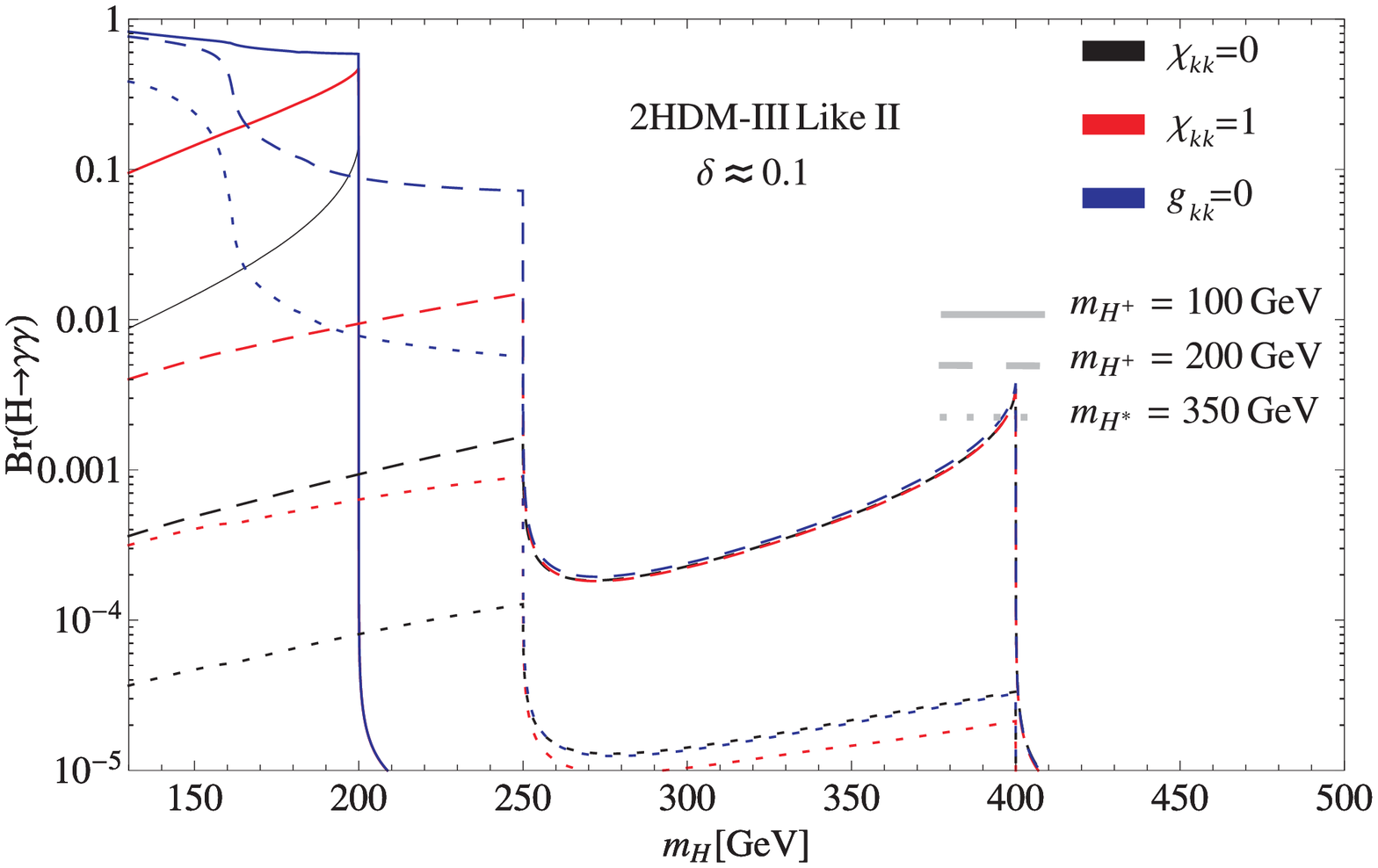},\includegraphics[width=2.9in]{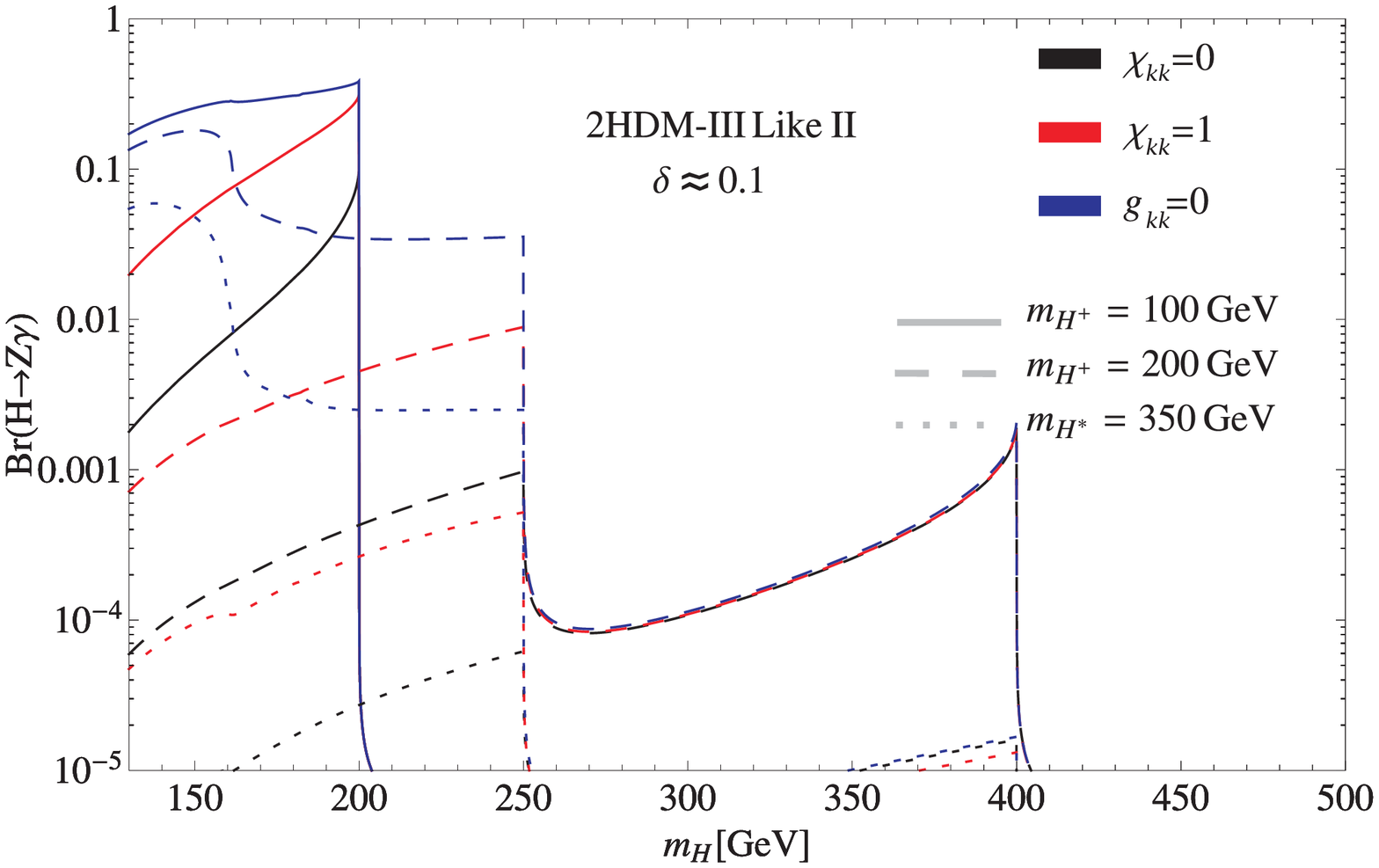}\\
  \caption{Decay rates  for the channels $H\to \gamma\gamma$ (left frame) and $H\to \gamma Z$ (right frame)
versus the heavy CP-even Higgs mass
for the
 2HDM-III Like II.
   The parameters used  here are as follows: $m_A=200$ GeV,  $\lambda_7=-\lambda_6=-1$, $\chi_{kk}^f=1$, $\chi_{23}^d=-0.35$,
 $\chi_{23}^u=-0.75$ and $X=10$. 
}
\label{H-decay}
\end{figure}

Next, we present the results for the  Br$(H\to\gamma\gamma)$ and Br$(H\to \gamma Z)$ versus the heavy Higgs boson  mass for three
different values of the charged Higgs boson one (see figure \ref{H-decay}). The scenarios presented in these plots are:
the fermiophobic one ($g_{kk}=0$), the one with Yukawa couplings mimicking a $\mathcal{Z}_2$ symmetry ($\chi_{kk}^f=0$) and a general 2HDM. Before the $H\to hh$ decay is
allowed, the differences between these scenarios are relevant, about two orders of magnitude (this between $g_{kk}=0$ and
$\chi_{kk}^f=0$ at $m_{H^+}=200$ GeV). However, when a light charged Higgs boson mass is considered ($m_{H^+}=100$ GeV), the
scenarios $\chi_{kk}^f=0$ and $\chi_{kk}^f=1$ yield similar rates for the $\gamma\gamma$ and $\gamma Z$
decay channels, both with Brs around $10^{-1}$. Finally, the Br$(H\to
\gamma\gamma)$ and Br$(H\to  \gamma Z)$ are disadvantaged when the heavy Higgs boson mass allows for the channels $H\to hh, AA$ or $H^+H^- $ to be open as, after this happens, these loop decays
are reduced to below the $10^{-5}$ level.

\subsection{The $A \to \gamma \gamma, \, \gamma Z $ decays}

\begin{figure}
  \centering
  \includegraphics[width=2.9in]{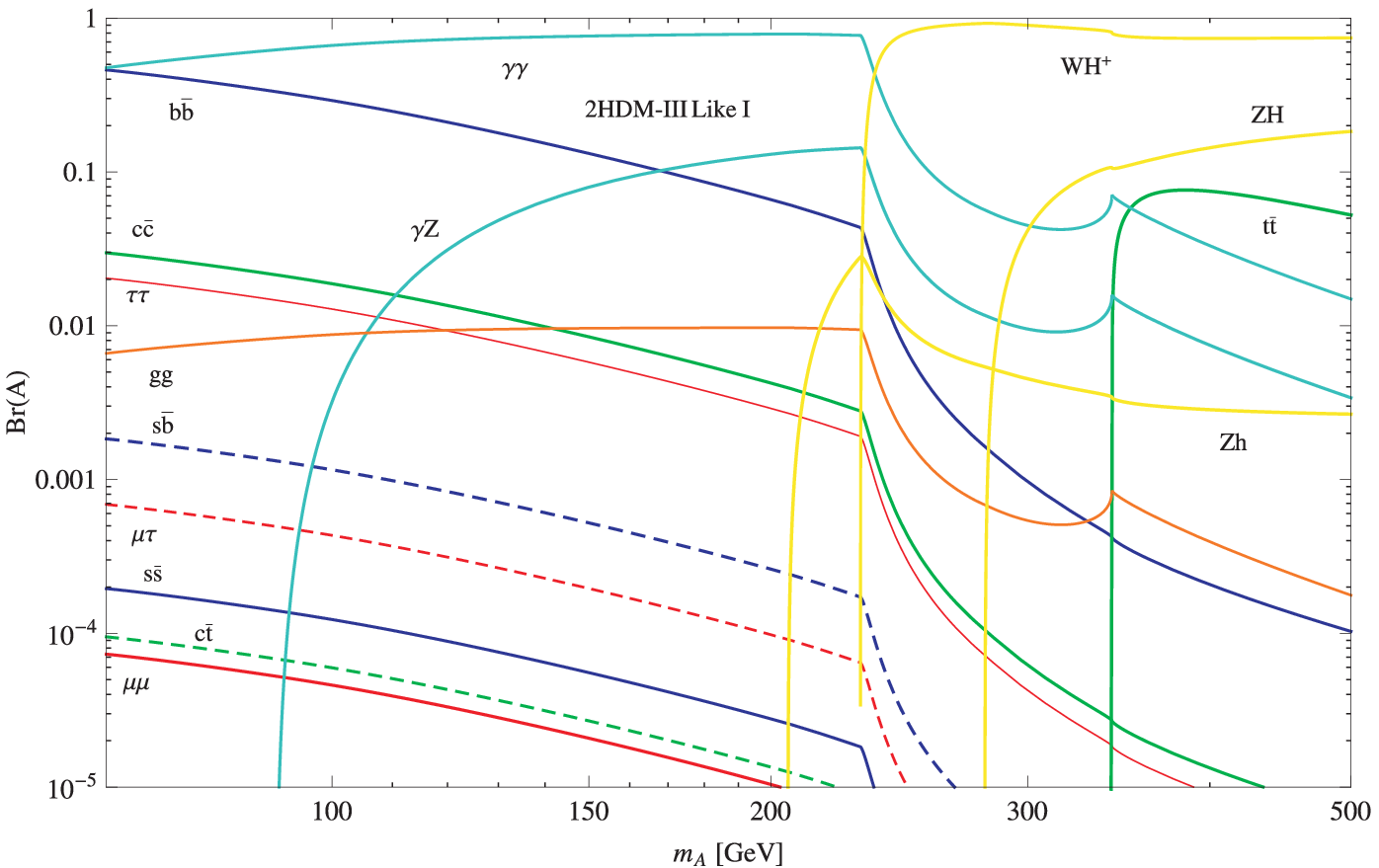}\includegraphics[width=2.9in]{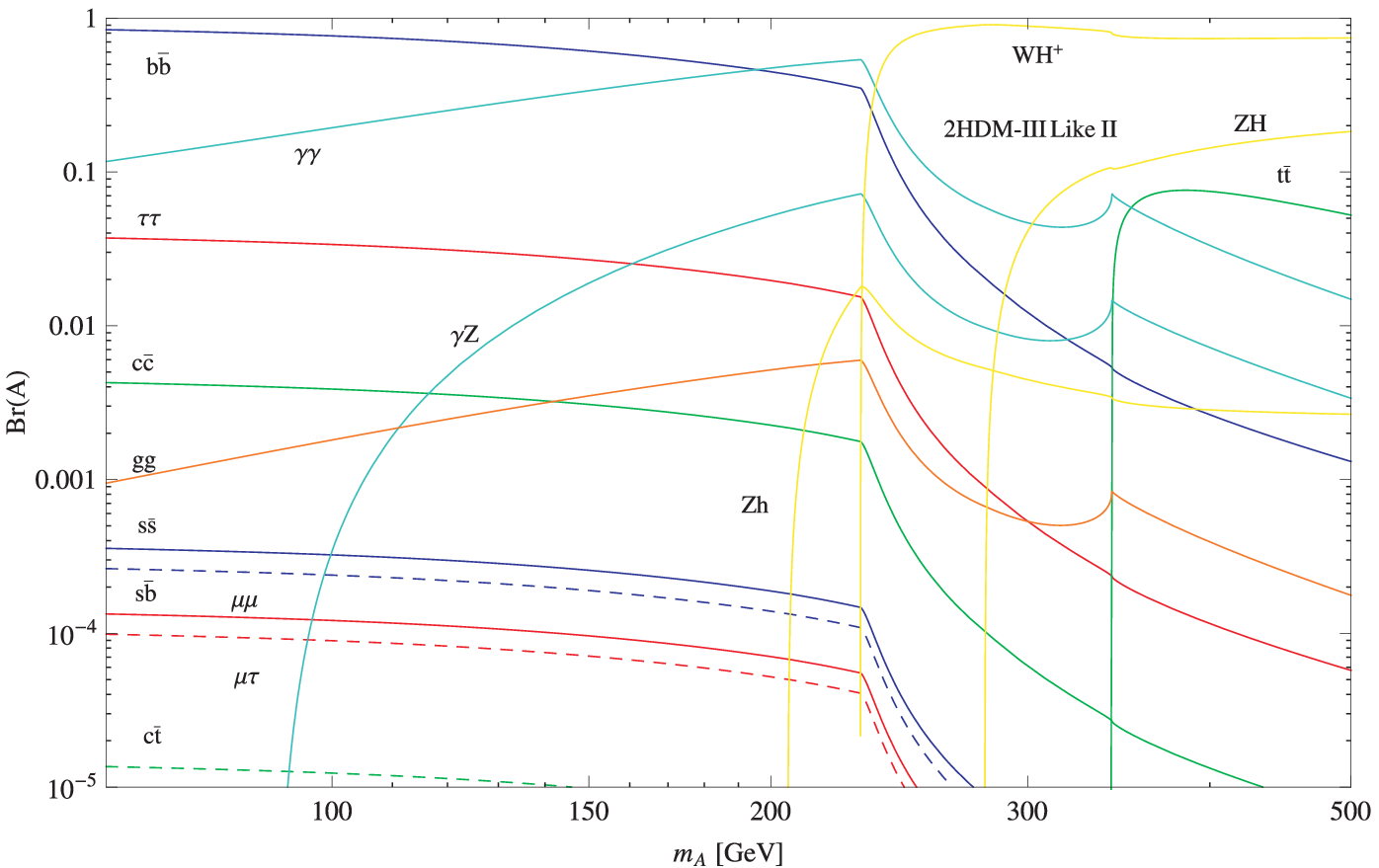}\\
  \caption{Behavior of all decay channels of the  CP-odd Higgs boson $A$ with respect to its mass for the
 2HDM-III-like I (left frame)
   and 2HDM-III-like II (right frame). The parameters used are as
follows:  $m_H=200$ GeV, $m_{H^+}=150$ GeV,
   $\lambda_7=-\lambda_6=-1$, $\chi_{kk}^f=1$, $\chi_{23}^d=-0.35$,
   $\chi_{23}^u=-0.75$ and $X=10$.
   }\label{all-A}
\end{figure}

In this last result subsection, we illustrate the decay phenomenology of the CP-odd Higgs boson, denoted by $A$. We start
our discussion with figure  \ref{all-A}, where we present the behavior of all decay channels  in the context of the
2HDM-III-like I and II. The coupling $Af\bar f$  presents high sensitivity to the underlying model, since while for the
2HDM-III-like I case it is proportional to $\cot\beta$ and for the 2HDM-III-like II it is proportional to $\tan\beta$. On the one hand, for the
case of the 2HDM-III-like I (left plot) the most relevant decay is $A\to\gamma \gamma$ as the $A\to b\bar b$ decay rate
is reduced by a factor of $1/10$ (as we have considered the choice $X=\tan\beta=10$). On the other hand, in the 2HDM-III-like II context (right plot)
the relevant decay is $A\to b\bar b$ for the opposite reason (for large $\tan\beta$). However, even in this last case the decay
$A\to \gamma\gamma$ presents a  size which is relevant, as Br$(A\to\gamma\gamma)\sim 10^{-1}$.
In general, when the $m_A$ value is large enough to allow for
the decays to $WH^+$, $Zh$ or $ZH$, the channels $A\to\gamma\gamma$ and $ \gamma Z$ are reduced by an order of magnitude. However, this pair of channels continue to be relevant.

\begin{figure}
  \centering
  \includegraphics[width=2.9in]{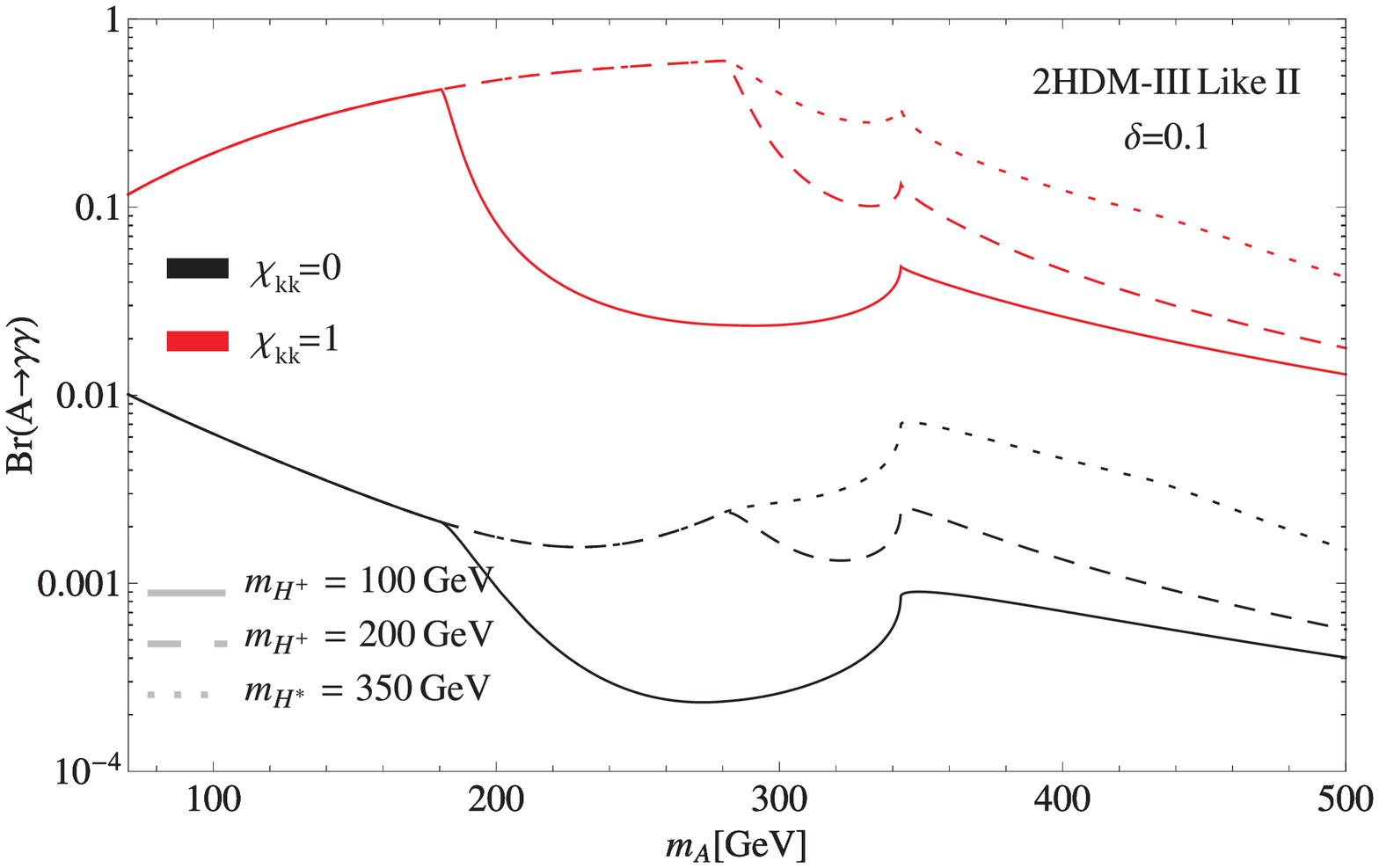}  \includegraphics[width=2.9in]{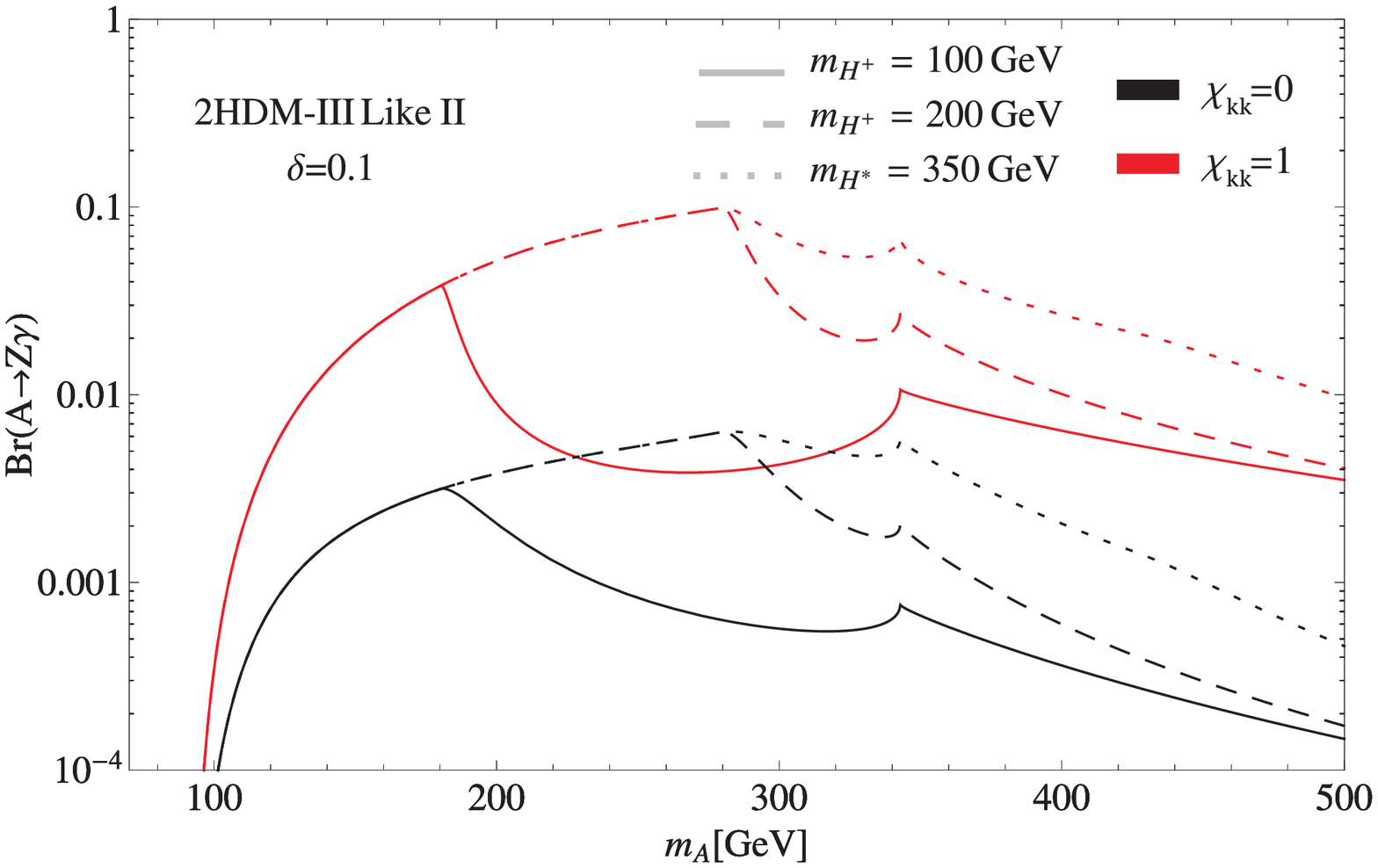}\\
  \caption{Decay rates  for the channels $A\to \gamma\gamma$ (left frame) and $A\to \gamma Z$ (right frame) versus the CP-odd Higgs mass
for the 2HDM-III-like II.
   The parameters used  here are as follows:
 $m_H=200\ GeV$,  $\lambda_7=-\lambda_6=-1$, $\chi_{kk}^f=1$, $\chi_{23}^d=-0.35$,
   $\chi_{23}^u=-0.75$ and $X=10$.
   }\label{A-decay}
\end{figure}

In figure \ref{A-decay} we present the Br$(A\to\gamma\gamma)$ and Br$(A\to\gamma Z)$ versus
the $A$ boson mass and for three different values of $m_{H^+}$. Unlike the previous CP-even states, for the case of
the CP-odd Higgs boson it  is
 impossible to implement a fermiophobic scenario, because only the fermionic particles contribute to the loops. For this reason,
 we can see that the difference between the two scenarios $\chi_{kk}^f=1$ and $\chi_{kk}^f=0$ can be up to two orders of magnitude.  The most relevant results
  are achieved via the $\chi_{kk}^f=1$ scenario, yielding a Br of ${\cal O}(10^{-1})$ for $\gamma\gamma$ and
of ${\cal O}(10^{-2})$
  for $\gamma Z$. In the same plots, again, it can be observed the strong sensitivity to the channels $A\to H^+W, hZ$
and $ HZ$ since, once
  these channels are open, the loop BRs decrease by about an order of magnitude.

\section{Conclusions}

We have studied  the significant enhancement
of the Brs of the decay channels  $h \to \gamma \gamma$ and $h\to \gamma Z$ in the context of the
2HDM-III,  assuming a four-zero Yukawa texture and a general Higgs potential.  We have shown that these processes are very sensitive to the flavor structure represented by such a Yukawa texture and to the triple Higgs couplings
entering the Lagrangian of the scalar sector. We also have shown that it is possible to accommodate the parameters of the model in such a way to obtain  the decay $h \to \gamma \gamma$ rates reported by the LHC and we have found a decay
rate for $h\to \gamma Z$ up to one order of magnitude larger than that one obtained in the SM, hence
amenable to experimental investigation with current and/or future LHC data.  We have then presented some benchmarks  where the parameters of the scenario considered are  consistent with all current experimental constraints.  In addition, we have found that it is possible to have a light charged Higgs boson compatible with all such measurements too, thereby
serving the purpose of being the smoking gun signal of the model considered, particularly in its Like II incarnation. We can finally confirm that the aforementioned loop decays can be enhanced, with respect to the corresponding SM rates, also for the
case of the heavy Higgs state $H$ while for the $A$ one (which has no SM counterpart). The corresponding rates can be sizable in certain regions of the 2HDM-III Like II parameter space for both Higgs states as well as the 2HDM-III Like I
for the latter only. We finally note that the scaling of the $\gamma\gamma$ and $\gamma Z$ rates with respect to the corresponding ones in the SM is not the same, unlike the case of many other BSM scenarios, thereby offering an alternative means to extract evidence of the most general 2HDM-III considered here.

\section*{Acknowledgements}
This work has been supported in part by\textit{ SNI-CONACYT (M\'exico)} and by \textit{PROMEP
(M\'exico)} under the grant ``Red Tem\'atica: F\'isica del Higgs y del sabor''.
SM is financed in part through the NExT Institute. He is also grateful for the hospitality extended to him
by the Benem\'erita Universidad Aut\'onoma de Puebla, where part of this work was carried out.

 \appendix

 \section{Higgs bosons tree level decays}

In this appendix we present explicitly the decay formulae for the neutral Higgs states of the 2HDM-III at tree level \cite{decays}. Notice that these have been written according
to the notation used in this work.

\subsection{Decay into  fermions pairs}

We first present the decay of a neutral Higgs boson to pairs of fermions,  without
FCNCs (like in the SM and the standard 2HDM with a $\mathcal{Z}_2$ symmetry). These decays can be written as follows:
\begin{equation}
\Gamma(\phi_i\to f\bar f)=\frac{N_c m_\phi}{8\pi}\left(\frac{g
m_f}{2m_W}\right)^2\left(1-\frac{4m_f^2}{m_\phi^2}\right)^{\rho/2}{\cal
G}_{\phi f\bar f}^2,
\end{equation}
where $\rho=3$ if $\phi=h,H$  and $\rho=1$ for $\phi=A$. However, in the
2HDM-III  it is indeed possible to have FCNCs,
so that it is important to know the corresponding decays, whichever their size, which are:
\begin{eqnarray}
\Gamma(\phi\to f_i\bar f_j)&=&\frac{N_c}{8\pi
m_{\phi}}\left(\frac{gm_i}{2m_W}\right)^2\big[m_{\phi}^2-(m_i+(-1)^{n}m_j)^2\big]\nonumber\\
&&\times\sqrt{\left[1-\left(\frac{m_i-m_j}{m_{\phi}}\right)^2\right]
\left[1-\left(\frac{m_i+m_j}{m_{\phi}}\right)^2\right]}{\cal
G}_{\phi f_i \bar f_j}^2,
\end{eqnarray}
here  $n=0$ for a Higgs boson which is CP-even and $n=1$ for the CP-odd state.

\subsection{Decay into vector particles}

One more possibility is that Higgs particles decay into two real gauge bosons. These decay channels can be written as
\begin{equation}
\Gamma(\phi_a\to VV)=\frac{G_f m_\phi^3}{16\sqrt{2}\pi}\delta_V\sqrt{1-4x}(1-4x+12x^2) {\cal G}_{\phi VV}^2,\ \ \ V=\{Z,W\},
\end{equation}
where $G_f$ is the Fermi constant, $x=m_V^2/m_\phi^2$ and $\delta_W=2$ and $\delta_Z=1$. Another option is to have one virtual
gauge boson, for this case the partial width is
\begin{eqnarray}
\Gamma(\phi_a\to VV^*)&=&\frac{3G_f^2M_V^4}{16\pi^3}m_\phi\delta'\left[\frac{3(1-8x+20x^2)}{(4x-1)^{1/2}}
arccos\left(\frac{3x-1}{2x^{3/2}}\right)\right.\nonumber\\
&&\left.-\frac{1-x}{2x}(2-13x+47x^2)-\frac{3}{2}(1-6x+4x^2)\log x\right]{\cal G}_{\phi VV}^2,
\end{eqnarray}
with $\delta_W'=1$ and $\delta_Z'=\frac{7}{12}-\frac{10}{9}s_W^2+\frac{40}{9}s_W^4$.

With respect to the CP-odd state, $A$, there are two channels:
\begin{equation}
\Gamma(A\to WH^+)=\frac{g[(m_W^2+m_{H^+}^2-m_A^2)-4m_{H^+}^2m_W^2]^{3/2}}{16m_A^3m_W^2\pi},
\end{equation}

\begin{equation}
\Gamma(A\to Z\phi)=\frac{g^2[(m_Z^2+m_\phi^2-m_A^2)^2-4m_\phi^2m_Z^2]^{3/2}}{64\pi m_Z^2m_A^3c_W^2}{\cal G}_\phi^2,
\end{equation}
with ${\cal G}_h=c_{\beta-\alpha}$, and ${\cal G}_H=s_{\beta-\alpha}$.

\subsection{Decay into gluons}

Now we present the decay into pairs of gluons. We begin with decays for the CP-even Higgs bosons:
\begin{equation}
\Gamma(\phi\to gg)=\frac{\alpha_s^2g^2m_\phi^2}{128\pi^3m_W^2}\left|\sum_q\tau_q[1+(1-\tau_q)f(\tau_q)]\right|^2g_{\phi ff}^2,
\end{equation}
where $\tau_q=4m_q^2/m_\phi^2$ and
\begin{equation}
  f_(\tau_q)=\left\{
  \begin{array}{ll}
  \arcsin(\sqrt{1/\tau_q})^2&if\ \tau_q\geq 1,\\
  \frac{1}{4}[\log(\eta_+/\eta_-)-i\pi]^2&if\ \tau_q<1,
  \end{array}
  \right.
\end{equation}
with $\eta_\pm=(1\pm\sqrt{1-\tau_q})$. For the CP-odd state,  $A$, we have instead:
\begin{equation}
  \Gamma(A\to gg)=\frac{\alpha_s^2g^2m_A^3}{128\pi^3m_W^2}\left|\sum_q\tau_qf(\tau_q)\right|^2{\cal G}_{Aff}^2.
\end{equation}

\subsection{Decay into  Higgs bosons }

Finally, the possibility of Higgs-to-Higgs decays is presented in this subsection.
We start with the widths for pairs of neutral Higgs bosons:
\begin{eqnarray}
\Gamma(\phi\to AA)&=&\frac{{\cal G}_{\phi AA}^2 }{32m_\phi^2\pi}\sqrt{m_\phi^2-4m_A^2},\\
\Gamma(H\to hh)&=&\frac{{\cal G}_{Hhh}^2 }{32m_H^2\pi}\sqrt{m_H^2-4m_h^2},
\end{eqnarray}
with
\begin{eqnarray}
{\cal G}_{hAA}&=&\frac{-g}{8m_W}\left\{8m_A^2s_{\beta-\alpha}+2m_h^2\frac{c_{\alpha-3\beta}+3c_{\beta+\alpha}}{s_{2\beta}}
-2m_{H^+}^2\left(\frac{c_\alpha}{c_\beta}-1\right)(s_{\alpha+3\beta}-3s_{\beta-\alpha})
\right.\nonumber\\
&&\left.
-16\mu_{12}^2\frac{c_{\beta+\alpha}}{s_{2\beta}^2} +\frac{8m_W^2c_{\beta-\alpha}}{g^2}\left(\frac{\lambda_6}{s_\beta^2}-\frac{\lambda_7}{c_\beta^2}\right)\right\},\\
{\cal G}_{HAA}&=&\frac{-g}{8m_W}\left\{8m_A^2c_{\beta-\alpha}+2m_H^2\frac{s_{\alpha-3\beta}+3s_{\beta+\alpha}}{s_{2\beta}}
+2m_{H^+}^2\left(\frac{c_\alpha}{c_\beta}-1\right)(c_{\alpha+3\beta}+3s_{\beta-\alpha})
\right.\nonumber\\
&&\left.
-16\mu_{12}^2\frac{s_{\beta+\alpha}}{s_{2\beta}^2} -\frac{8m_W^2s_{\beta-\alpha}}{g^2}\left(\frac{\lambda_6}{s_\beta^2}-\frac{\lambda_7}{c_\beta^2}\right)\right\},\\
{\cal G}_{Hhh}&=&\frac{-gc_{\beta-\alpha}}{2m_Ws_{2\beta}^2}\left\{(2m_h^2+m_H^2)s_{2\alpha}s_{2\beta}
+2\mu_{12}^2(s_{2\beta}-3s_{2\alpha})\right.\nonumber\\
&&\left.+\frac{12 m_W^2s_{2(\alpha-\beta)}}{g^2}(\lambda_6c_\beta^2-\lambda_7s_\beta^2)\right\}.
\end{eqnarray}

Finally, the width for $H\to H^+H^-$ can be written as follows:
\begin{equation}
  \Gamma(H\to H^+H^-)=\frac{g^2{\cal G}_{HH^+H^-}^2m_W^2}{256m_H^2\pi}\sqrt{m_H^2-4m_{H^+}^2}.
\end{equation}

 \section{Couplings}

In this section we present all the couplings that we have used for this work. These will be presented in
general form, together with the explicit factors for every scenario.

\subsection{Fermion couplings}

We begin with the Yukawa couplings, which have already appeared in section  II of this work.
In a general way, these couplings are:
\begin{eqnarray}
g_{\phi_a f f}&=&\frac{-ig m_f}{2m_W} {\cal G}_{\phi_a f
f},\\
g_{A f f}&=& \frac{g m_f}{2 m_W} \gamma^5 {\cal G}_{A f
f},
\end{eqnarray}
where the factors ${\cal G}$ are defined in table \ref{Yukawa-table}.
\begin{table}[h]
\begin{center}
\begin{tabular}{|c|c|c|c|}
  \hline
  ${\cal G}_{\phi_i  ff}$ & leptons& quarks-down & quarks-up \\
  \hline
  $h$ & $\xi_h^l\delta_{ij}+\frac{\xi_H^l-Z\xi_h^l}{\sqrt{2}f(Z)}\sqrt{\frac{m_{l_j}}{m_{l_i}}}\chi_{ij}^l$
  & $\xi_h^d\delta_{ij}+\frac{\xi_H^d-X\xi_h^d}{\sqrt{2}f(X)}\sqrt{\frac{m_{d_j}}{m_{d_i}}}\chi_{ij}^d$
  &$\xi_h^u\delta_{ij}-\frac{\xi_H^u+Y\xi_h^u}{\sqrt{2}f(Y)}\sqrt{\frac{m_{u_j}}{m_{u_i}}}\chi_{ij}^u$ \\
  \hline
  $H$ & $\xi_H^l\delta_{ij}-\frac{\xi_h^l-Z\xi_H^l}{\sqrt{2}f(Z)}\sqrt{\frac{m_{l_j}}{m_{l_i}}}\chi_{ij}^l$
  &$\xi_H^d\delta_{ij}-\frac{\xi_h^d-X\xi_H^d}{\sqrt{2}f(X)}\sqrt{\frac{m_{d_j}}{m_{d_i}}}\chi_{ij}^d$
  &$\xi_H^u\delta_{ij}+\frac{\xi_h^u+Y\xi_H^u}{\sqrt{2}f(Y)}\sqrt{\frac{m_{u_j}}{m_{u_i}}}\chi_{ij}^u$ \\
  \hline
  $A$ & $-Z\delta_{ij}+\frac{f(Z)}{\sqrt{2}}\sqrt{\frac{m_{l_j}}{m_{l_i
}}}\chi_{ij}^l$& $-X\delta_{ij}+\frac{f(X)}{\sqrt{2}}\sqrt{\frac{m_{d_j}}{m_{d_i
}}}\chi_{ij}^d$&$-Y\delta_{ij}+\frac{f(Y)}{\sqrt{2}}\sqrt{\frac{m_{u_j}}{m_{u_i
}}}\chi_{ij}^u$\\
  \hline
\end{tabular}
\end{center}
\caption{Dimensionless functions that define the Yukawa
couplings in the 2HDM-III.}\label{Yukawa-table}
\end{table}

Others couplings needed for this work are the vector-fermion-fermion couplings. Essentially, for this analysis we
 need $\gamma f\bar f$ and $Z f\bar f$, which are described as
\begin{eqnarray}
g_{\gamma  ff}&=&-ieQ_f\gamma_\mu,\\
g_{Z f f}&=&\frac{ig}{4c_W}\gamma_\mu(F_V-F_A\gamma_5),
\end{eqnarray}
where, $F_V$($F_A$) represents the vectorial(axial) part of the couplings and their explicit form is shown in table
\ref{Axial-verctor}.

\begin{table}[ht]
  \centering
  \begin{tabular}{|c|c|c|c|}
    \hline
     & for $u$-quarks & for $d$-quarks & for leptons \\
    \hline
    $F_V$ & $1-\frac{8}{3}s_W^2$ & $1+\frac{4}{3}s_W^2$ & $-1+4s_W^2$ \\
    \hline
    $F_A$ & $-1$ &$ 1$ & $1$ \\
    \hline
  \end{tabular}
  \caption{Axial and vector components for the $Z\bar f f$ couplings.}\label{Axial-verctor}
\end{table}

\subsection{Gauge sector}
Now, we write the couplings for the gauge sector. For this calculation it is convenient to adopt the unitary gauge, so that the couplings
 $V^\mu(k_1)W^{+\lambda}(k_2)W^{-\rho}(k_3)$ and $V_1^\mu V_2^\nu
W^{+\lambda} W^{-\rho}$ (where the $V^\mu$s represent any neutral vector boson) can be written as
$ig_V\Gamma_{\lambda\rho\mu(k_1,k_2,k_3)}$ and
$ig_{V_1}g_{V_2}\Gamma_{\lambda\rho\mu\nu}$, with
\begin{eqnarray}
\Gamma_{\lambda\rho\mu}(k_1,k_2,k_3)&=&(k_2-k_3)_\mu
g_{\lambda\rho}+\Big(k_3-k_1\Big)_\lambda
g_{\rho\mu}+\Big(k_1-k_2\Big)_\rho
g_{\lambda\mu},\\
\Gamma_{\lambda\rho\mu\nu}&=&-2g_{\mu\nu}g_{\lambda\rho}+g_{\lambda\mu}g_{\rho\nu}+g_{\rho\mu}g_{\lambda\nu},
\end{eqnarray}
$g_\gamma=gs_W$ and $g_Z=gc_W$.
\subsection{Scalar and kinetic sector}

\begin{table}[!h]
  \centering
\begin{tabular}{|c|c|c|c|}
  \hline
  Coupling & Vertex Function & Coupling & Vertex Function \\
  \hline
  $g_{\phi_i H^\pm H^\mp}$ & $\frac{-igm_W}{4}{\cal G}_{\phi_iH^\pm H^\mp}$ &$g_{\phi_a WW}$&$ig m_W{\cal G}_{\phi_aWW}g_{\mu\nu}$\\
  \hline
  $g_{\gamma H^\pm H^\mp}$ & $i e(P_--P_+)_\mu$&$g_{ZH^\pm H^\mp}$&$ie t_{2W}^{-1}(P_--P_+)_\mu$\\
  \hline
  $g_{\gamma\gamma H^\pm H^\mp}$&$2ie^2 g_{\mu\nu}$&$g_{Z\gamma H^\pm H^\mp}$&$2ie^2t_{2W}^{-1}g_{\mu\nu}$\\
  \hline
\end{tabular}
  \caption{The three- and four-particle couplings between scalars and vectors in the 2HDM-III.}\label{scalar-couplings}
\end{table}
Finally, the couplings between scalar particles themselves and scalar-vector-vector couplings are presented in table \ref{scalar-couplings}.
Herein, we have ${\cal G}_{hWW}=s_{\beta-\alpha}$ (${\cal G}_{HWW}=c_{\beta-\alpha}$) and
\begin{eqnarray}
{\cal G}_{h H^+H^-}&=&\frac{-1}{16 g^2 m_W^2}\Big\{16 g^2\mu_{12}^2\frac{c_{\alpha+\beta}}{s_{2\beta}^2}+\frac{g^2m_{H^+}^2}{c_\beta}\Big(s_{\alpha-2 \beta}+3 s_{2 \alpha-\beta}-s_{\alpha+2\beta}+s_{2 \alpha+3 \beta}\nonumber\\
&&-s_{\alpha+4\beta}+s_{\alpha}-3 s_{\beta}+s_{3 \beta}\Big)-2g^2m_h^2\frac{c_{\alpha-3\beta}+3c_{\alpha+\beta}}{s_{2\beta}}- 8m_W^2\lambda_6\frac{c_{\alpha-\beta}}{s_{\beta}^2}\nonumber\\
&&+8m_W^2\lambda_7\frac{c_{\alpha-\beta}}{c_\beta^2}\Big\},\\
{\cal G}_{H H^+H^-}&=&\frac{-1}{16 g^2 m_W^2}\Big\{16 g^2\mu_{12}^2\frac{s_{\alpha+\beta}}{s_{2\beta}^2}-\frac{g^2m_{H^+}^2}{c_\beta}\Big(c_{\alpha-2 \beta}+3 c_{2 \alpha-\beta}-c_{\alpha+2\beta}+c_{2 \alpha+3 \beta}\nonumber\\
&&-c_{\alpha+4\beta}+c_{\alpha}+3 c_{\beta}+c_{3 \beta}\Big)-2g^2m_H^2\frac{s_{\alpha-3\beta}+3s_{\alpha+\beta}}{s_{2\beta}}- 8m_W^2\lambda_6\frac{s_{\alpha-\beta}}{s_{\beta}^2}\nonumber\\
&&+8m_W^2\lambda_7\frac{s_{\alpha-\beta}}{c_\beta^2}\Big\}.
\end{eqnarray}

\end{document}